\documentclass[11pt]{article}
\usepackage[left=0.8in,top=0.5in,right=0.8in,bottom=1.5in,head=.1in,nofoot]{geometry}

\usepackage{setspace,url,bm,amsmath,amsfonts,dsfont} 
\usepackage{makecell}
\usepackage{titlesec} 
\titlelabel{\thetitle.\quad} 

\usepackage[margin=20pt]{subcaption}
\RequirePackage{amsthm,amsmath,amsfonts,amssymb}
\RequirePackage[authoryear]{natbib}
\RequirePackage[colorlinks,citecolor=blue,urlcolor=blue]{hyperref}
\RequirePackage{graphicx}
\RequirePackage{mathrsfs}  
\RequirePackage{bm}
\RequirePackage{graphicx}
\RequirePackage{graphics}
\RequirePackage{multicol}
\RequirePackage{multirow}
\RequirePackage{enumitem}
\RequirePackage{booktabs}
\RequirePackage[table,xcdraw]{xcolor}
\usepackage{indentfirst}
\usepackage{authblk}

\usepackage{appendix}
\usepackage[plain,shortend,linesnumbered,ruled,lined]{algorithm2e}
\theoremstyle{plain}
\newtheorem{theorem}{Theorem}
\newtheorem{corollary}{Corollary}
\newtheorem{lemma}{Lemma}

\theoremstyle{remark}
\newtheorem{example}{Example}

\newtheorem{assumption}{Assumption}
\usepackage{arydshln}
\usepackage{xcolor}
\usepackage{comment}
\usepackage{colortbl}
\usepackage{pifont}
\usepackage{placeins}

\newcommand{\vt}{{\vartheta}}
\newcommand{\bmvt}{{\bm \vt}}

\def\T{{ \mathrm{\scriptscriptstyle T} }}

\newcommand{\bbP}{{\mathbb P}}
\newcommand{\bbR}{{\mathbb R}}

\newcommand{\bbU}{{\mathbb U}}
\newcommand{\bbE}{{\mathbb E}}

\newcommand{\bz}{{\boldsymbol z}}

\newcommand{\bE}{{\boldsymbol E}}
\newcommand{\bF}{{\boldsymbol F}}
\newcommand{\bM}{{\boldsymbol M}}
\newcommand{\bI}{{\boldsymbol I}}

\newcommand{\bX}{{\boldsymbol X}}
\newcommand{\bY}{{\boldsymbol Y}}

\newcommand{\bW}{{\boldsymbol W}}

\newcommand{\bZ}{{\boldsymbol Z}}

\newcommand{\bgamma}{{\boldsymbol \gamma}}

\newcommand{\btheta}{{\boldsymbol \theta}}

\newcommand{\cE}{{\mathcal E}}
\newcommand{\cF}{{\mathcal F}}

\newcommand{\cN}{{\mathcal N}}
\newcommand{\cP}{{\mathcal P}}

\newcommand{\cT}{{\mathcal T}}

\newcommand{\cZ}{{\mathcal Z}}

\def\x{\boldsymbol{x}}
\def\star{\obs}
\def\obs{{\rm{obs}}}
\def\mis{{\rm{mis}}}
\def\imp{{\rm{imp}}}
\def\p{{\rm{p}}}
\def\e{{\rm{e}}}
\def\s{{\rm{s}}}
\def\o{{\rm{o}}}
\def\b{{\rm{b}}}
\def\m{{\rm{m}}}
\def\db{{\rm{db}}}
\def\pb{{\rm{pb}}}

\def\SI{{\rm{SI}}}
\def\MI{{\rm{MI}}}

\def\unif{{\rm{Unif}}}

\begin{document}

\title{\bf Imputation-based randomization tests for randomized experiments with interference}
\author[1]{Tingxuan 
Han$^{*}$}
\author[2,3]{Ke Zhu\footnote{These authors have contributed equally to this work.}}
\author[1]{Hanzhong Liu$^\dagger$}
\author[1]{Ke Deng\footnote{Co-corresponding authors: lhz2016@tsinghua.edu.cn; kdeng@tsinghua.edu.cn}}
\affil[1]{Department of Statistics and Data Science, Tsinghua University}
\affil[2]{Department of Statistics, North Carolina State University}
\affil[3]{Department of Biostatistics and Bioinformatics, Duke University}
\date{}
\renewcommand\Authands{ and }

\maketitle

\begin{abstract}
The presence of interference renders classic Fisher randomization tests infeasible due to nuisance unknowns. 
To address this issue, we propose imputing the nuisance unknowns and computing Fisher randomization \textit{p}-values multiple times, then averaging them. 
We term this approach the imputation-based randomization test and provide theoretical results on its asymptotic validity. 
Our method leverages the merits of randomization and the flexibility of the Bayesian framework: for multiple imputations, we can either employ the empirical distribution of observed outcomes to achieve robustness against model mis-specification or utilize a parametric model to incorporate prior information.
Simulation results demonstrate that our method effectively controls the type I error rate and significantly enhances the testing power compared to existing randomization tests for randomized experiments with interference.
We apply our method to a two-round randomized experiment with multiple treatments and one-way interference, where existing randomization tests exhibit limited power.
\end{abstract}

\section{Introduction}\label{sec:intro}

In causal inference, the ``no interference" assumption posits that potential outcomes of each unit are unaffected by assignments of other units \citep{cox1958planning,rubin1980randomization}. 
In practice, however, interference does exist in different scenarios. 
For instance, whether or not other people were vaccinated against COVID-19 can affect your probability of infection. Causal inference in the presence of interference has received much attention in recent years \citep{rosenbaum2007interference,hudgens2008toward,tchetgen2012causal,liu2014large,loh2020randomization,imai2021causal,yu2022estimating,leung2022causal,forastiere2022estimating,vazquez2023identification,shirani2024causal}. 

As the golden standard for drawing causal conclusions, randomized experiments provide a solid basis for causal inference and enable \emph{Fisher randomization tests} (FRTs) exactly valid in finite samples under the assumption of no interference \citep{fisher1935design}. 
In the presence of interference, however, classic FRTs are infeasible since the null hypothesis of interest can no longer impute all potential outcomes certainly.
To address this challenge, \emph{conditional randomization tests} (CRTs) have been proposed to conduct causal inference for a selected subset of units (referred to as the focal units) with respect to a carefully selected subset of treatment assignments (referred to as the focal assignments) instead, so that the null hypothesis is sharp for the conditional cause inference and thus the classic FRT is still applicable \citep{aronow2012general,athey2018exact,basse2019randomization,puelz2022graph,yanagi2022improving,owusu2023randomization,tiwari2024quasi,basse2024randomization,zhang2024multiple}. 
Although CRTs can achieve valid causal inference, they often suffer from severe power loss due to aggressive selection of focal units and assignments in conditioning \citep{puelz2022graph}.
{\cite{zhong2024unconditional} proposed \textit{partial null randomization tests} (PNRTs), which avoid conditioning on focal assignments by conducting pairwise comparisons of the test statistic between the observed assignment and all possible assignments.
This method allows distinct focal unit sets for different assignments, enhancing statistical power. 
While PNRTs relax the restrictions of CRT, they remain underpowered in challenging cases where the focal unit proportion is low, as demonstrated in our simulations and real data application.}

Alternatively, \emph{multiple imputation} (MI) pioneered by
{\cite{rubin1977formalizing,rubin1978multiple,rubin1996multiple,rubin2004design,rubin2004multiple}}
is another popular approach to handle nuisance unknowns in causal inference, following the Bayesian idea.
Examples include imputing compliance status to enable more powerful FRTs with noncompliance \citep{rubin1998more,lee2017more,forastiere2018posterior}, imputing potential outcomes in FRTs for facilitating partially post hoc subgroup analyses \citep{lee2015valid}, handling unspecified factorial effects \citep{espinosa2016bayesian}, addressing missing outcomes \citep{ivanova2022randomization}, and managing disruptions (e.g., COVID-19) in clinical trials \citep{uschner2023using}.
{Noticeably, FRT with MI often leads to a posterior predictive \textit{p}-value \citep{rubin1984bayesianly}, enabling us to address more complex scenarios in causal inference from the Bayesian perspective} \citep{ding2023posterior, leavitt2023randomization, li2023bayesian,ray2020semiparametric,breunig2022double}.

{In this paper, we follow such an imputation-based idea for implementing FRTs to propose an \emph{imputation-based randomization test} (IRT) for randomized experiments with interference.
Imputing unobserved potential outcomes that cannot be fully specified based on the null hypothesis of interest with respect to an appropriate imputation distribution, we come up with a series of FRT \textit{p}-values, each of which comes from a FRT corresponding to a specific imputation of the unobserved potential outcomes.
A working \textit{p}-value, referred to as the IRT \textit{p}-value, can be obtained by averaging these FRT \textit{p}-values to claim statistical significance of the causal effect of interest.
Retaining all units and potential treatment assignments, the proposed IRT is more powerful than CRTs and PNRTs, and can be easily applied to various challenging scenarios with interference.
Theoretical analyses show that the proposed IRT controls the type I error rate at most twice the significance level asymptotically when the imputation distribution is properly specified.
Furthermore, numerical results show that IRT empirically maintains the type I error rate at the nominal significance level and substantially improves power over existing methods.
}

\section{Preliminaries}\label{sec:overview}
\subsection{The Potential Outcome Model in Presence of Interference}
Consider a finite population $\bbU = \{1,\ldots,N\}$ with $N$ units, to each of which $K$ different treatments in the treatment set $\cT=\{t_1,\ldots,t_K\}$ can be potentially assigned. 
For simplicity, we mainly consider the simple case with $K=2$ in this study, though our approach can be easily extended to more complicated scenarios where $K>2$ (as demonstrated in the real data application in Section \ref{sec:realdata}).
Let $\cT=\{0,1\}$ be the binary treatment set with $0$ as control treatment and 1 as active treatment, and $Z_i \in \cT$ be the indicator of treatment assignment for the $i$-th unit in $\bbU$. 
The collective treatment assignment for all $N$ units is represented by the binary vector $\bZ = (Z_1,\ldots,Z_N) \in \cZ \subseteq \{0,1\}^N$. 
Let $\bZ^{\obs} = (Z_1^\obs,\ldots,Z_N^\obs) \in\cZ$ be the treatment assignment actually implemented in the randomized experiment.
Throughout this study, we always assume that $\bZ^\obs$ is a random sample from a known randomized design $\bF_\cZ$ defined over $\cZ$.

In the presence of interference, each unit $i$ has a collection of potential outcomes, i.e., $\{Y_i(\bz)\}_{\bz\in\cZ}$,
in general, because there could be a total of $2^N$ distinct exposures for each unit in the most general case where interference occurs between any two units. 
In practice, however, interference typically occurs only among certain units.
In this case, the number of distinct exposures for each unit would be much smaller than $2^N$. 
In practice, researchers often assume that only a small exposure set $\cE$ is available, with the actual exposure received by unit $i$ under treatment assignment vector $\bZ$ determined by an exposure mapping $m_i: \cZ \rightarrow \cE$. 
This leads to the assumption of \emph{correct exposure mapping} as outlined below \citep{manski2013iden,aronow2017estimating,hoshino2023randomization}.
Apparently, the potential outcomes of unit $i$ reduce to a much smaller set $\{Y_i(a)\}_{a\in\cE}$ under such an assumption.

\begin{assumption}[Correct exposure mapping] \label{assump:correct_exposure}
For all $i \in \bbU$ and $\bz,\bz' \in \cZ$, if $m_i(\bz) = m_i(\bz')$, then we have $Y_i(\bz) = Y_i(\bz')$.
\end{assumption}

Different exposure mappings define different interference mechanisms.
A general framework for establishing an interference mechanism is to introduce an interference network $\cN$ over units of interest where units $i$ and $j$ are connected (denoted as $i\sim j$) if and only if interference may occur between them, and define 
\begin{equation}\label{eq:ExposureMapping}
m_i(\bz) = \begin{cases}
    2, & \text{if } z_i=1;\\
    1, & \text{if } z_i=0 \text{ and } z_j=1, \text{ for some unit }j\sim i;\\
    0, & \text{if } z_i=0 \text{ and } z_j=0, \text{ for any unit }j\sim i.
\end{cases}
\end{equation}
Under such an interference mechanism, we have three types of treatment exposures for each unit $i$, i.e., $\cE=\{0,1,2\}$, where exposure $2$ implies that unit $i$ itself is assigned to treatment, exposure $1$ implies that units $i$ is assigned to control while one of its direct neighbors is under treatment, and exposure $0$ implies that unit $i$ and all its direct neighbors are all under control. 
Many popular interference mechanisms can be viewed as special cases of this framework.
Below are three concrete examples.

\begin{example}[Clustered Interference]\label{eg:clustered}
Assuming that the $N$ units in the finite population of interest are grouped into $K$ disjointed clusters (such as villages), and interference occurs solely within clusters, \cite{basse2019randomization} considered an interference mechanism known as \emph{clustered interference} where $i\sim j$ if and only if units $i$ and $j$ belong to the same cluster.
\end{example}

\begin{example}[Spatial Interference]\label{eg:spatial}
Assuming that the $N$ units of interest are located in a spatial space with $d(i,j)$ being the spatial distance between units $i$ and $j$, and interference occurs only between two units whose spatial distance is shorter than a threshold $r$, \cite{puelz2022graph} considered an interference mechanism known as \emph{spatial interference} where $i\sim j$ if and only if $d(i,j)\leq r$.
\end{example}

\begin{example}[No Interference]\label{eg:NoInterference}
In case where there does not exist $j \not= i$ such that $i\sim j$, $m_i(\bz)$ in Eq. \eqref{eq:ExposureMapping} degenerates to $m_i(\bz) = 2z_i$ for all $i\in\bbU$ with $\cE=\{0,2\}$, which can be further simplified to $m_i(\bz) = z_i$ and $\cE = \{0,1\}$ after a rescaling of $z_i$'s. 
In this special case, the exposure unit $i$ receives is completely determined by $Z_i$, implying no-interference between units. 
This setting is a critical component of the \emph{stable unit treatment value assumption} (SUTVA) proposed by \cite{rubin1980randomization}. 
\end{example}

\subsection{Fisher Randomization Test}
In case of no interference, a classic problem in a randomized case-control study is to test the sharp null hypothesis of no treatment effect for all units. 
Following the notation in Example \ref{eg:NoInterference},
\begin{equation}\label{eq:FRT_NullHypothesis}
H_{0}^\cF: Y_i(1)=Y_i(0),\ 1\leq i\leq N,
\end{equation}
where $Y_i(1)$ and $Y_i(0)$ denote the potential outcomes of unit $i$ under treatment and control, respectively.
Under $H_{0}^\cF$, define $Y_i(1)=Y_i(0)=\theta_i$ for all $i\in\bbU$ and $\btheta = (\theta_1,\cdots,\theta_N)$. 
With the observed outcomes $\bY^{\obs}=\{Y_i^{\obs}\}_{i\in\bbU}$ where $Y_i^{\obs}=Z_i^\obs Y_i(1)+(1-Z_i^\obs)Y_i(0)=\theta_i$, $\btheta$ is fully observed.
Consider a finite-population framework where potential outcomes are fixed quantities, and the only source of randomness is the treatment assignment.
The classic difference-in-means test statistic under a treatment assignment $\bz$ is 
\begin{equation}\label{eq:FRT_TestStat}
T(\bz,\btheta) = \left|\frac{\sum_{i=1}^N z_i\theta_i}{\sum_{i=1}^N z_i}
- \frac{\sum_{i=1}^N (1-z_i)\theta_i}{\sum_{i=1}^N (1-z_i)}\right|.
\end{equation}
Because $\btheta$ is a fixed vector after the randomization experiment is conducted, the null distribution of $T(\bZ,\btheta)$ is determined by the randomization distribution {$\bF_\cZ$}.
Based on these facts, \emph{Fisher randomization test} (FRT) can be conducted by calculating the \textit{p}-value below:
\begin{equation}\label{eq:FRT_pValue}
p(\bZ^{\star};\btheta) = \bbP_\bZ\left(T(\bZ,\btheta) \geq T(\bZ^{\star},\btheta)\right),
\end{equation}
where $\bZ^{\star}$ stands for the realized assignment, and $\bbP_\bZ$ is the probability defined by 
$\bZ \sim \bF_\cZ$. 
According to \cite{fisher1935design}, FRT is finite-sample exact in the sense that under $H_{0}^\cF$, 
$$
\bbP\left(p(\bZ^{\star};\btheta) \leq \alpha\right) \leq \alpha\ \text{for all }0 \leq \alpha \leq 1,\text{ when }\bZ^{\star} \sim {\bF_\cZ}.
$$
Hypothesis $H_{0}^\cF$ is commonly referred to as the \emph{sharp null hypothesis}, since 
the reference distribution of the test statistic $T(\bZ,\btheta)$ is fully determined under $H_{0}^\cF$.

\subsection{Conditional and Partial Null Randomization Tests}
 
In presence of interference, however, the situation becomes more complicated.
Given the treatment exposure mapping in Eq. \eqref{eq:ExposureMapping}, we aim to test the null hypothesis on the contrast between any two exposures $a$ and $b$ in $\cE$, i.e.,
\begin{equation}\label{eq:contrast}
{H_0^{\{a,b\}}: Y_i(a) = Y_i(b)},\ 1\leq i\leq N.
\end{equation}
For example, 
we are often interested in testing the null hypothesis of no spillover effect, i.e., $H_0^{\{0,1\}}$.
Under $H_{0}^{\{a,b\}}$, define $Y_i(a)=Y_i(b)=\theta_i$ for all $i\in\bbU$ and $\btheta=(\theta_1,\cdots,\theta_N)$, then the classic difference-in-means test statistic under treatment assignment $\bz$ becomes
\begin{equation}\label{eq:CRT_TestStat}
T_{a,b}(\bz,\btheta) = \left|\frac{1}{N_b(\bz)}\sum_{i\in\bbU_b(\bz)} \theta_i
- \frac{1}{N_a(\bz)}\sum_{i\in\bbU_a(\bz)} \theta_i\right|,
\end{equation}
where $\bbU_b(\bz)=\{i\in\bbU:m_i(\bz)=b\}$, $N_b(\bz)=|\bbU_b(\bz)|$, $\bbU_a(\bz)=\{i\in\bbU:m_i(\bz)=a\}$, and $N_a(\bz)=|\bbU_a(\bz)|$.
$T_{a,b}(\bz,\btheta)$ is well defined as long as $N_a(\bz),N_b(\bz)>0$ for all $\bz \in \cZ$.

A critical challenge in testing $H_{0}^{\{a,b\}}$ in a randomized experiment with interference, however, lies in the fact that even under $H_{0}^{\{a,b\}}$, not all elements of $\btheta$ are observable after the experiment is conducted with the realized assignment $\bZ^{\star}$.
{To be concrete, define $\bbU^{\star}_{a,b}=\bbU_a(\bZ^{\star})\cup\bbU_b(\bZ^{\star})$, the partially observed version of $\btheta$ can be expressed as
\begin{equation}
\label{eq:thetaobs}
\btheta^\obs=(\theta^\obs_1,\ldots,\theta_N^\obs)\ \mbox{with}\ \theta^\obs_i=\theta_i\ \mbox{for}\ i\in\bbU_{a,b}^\obs\ \mbox{and}\ \theta^\obs_i=?\ \mbox{for}\ i\notin\bbU_{a,b}^\obs,
\end{equation}
where the question mark `$?$' stands for the missing values.
Define 
$N^{\star}_{a,b}=|\bbU^{\star}_{a,b}|$.
Because only $N^{\star}_{a,b}$ elements of $\btheta$ are observable, the null hypothesis $H_0^{\{a,b\}}$ is no longer sharp, resulting in an unknown reference distribution of the test statistic $T_{a,b}(\bZ,\btheta)$ under $H_0^{\{a,b\}}$.
}

To deal with this critical issue, \cite{aronow2012general} and \cite{athey2018exact} proposed \emph{conditional randomization tests} (CRTs) to substitute the classic FRT that is not applicable in this setting. 
The main idea is to avoid nuisance unknowns by
focusing only on a subset of carefully selected units, referred to as ``focal units", and conducting randomized experiments on them with a randomization distribution restricted to a subset of treatment assignments, referred to as ``focal assignments". This approach ensures that the null hypothesis $H_0^{\{a,b\}}$ becomes sharp in the conditional inference.
Because the reference distribution of the test statistic with focal units is precisely known conditional on the focal assignments and under the null hypothesis, an FRT-like testing procedure can be conducted.
\citet{basse2019randomization} has shown that CRTs are exact both conditionally and unconditionally under mild conditions. 
Nonetheless, because CRTs employ only a restrictive subset of treatment assignments on a small subset of units of interest, they often suffer from severe power loss.
To mitigate power loss, it is advisable to utilize as large a set of focal units associated with a set of focal assignments as possible, because a larger focal unit set often ensures a higher power, whereas a larger focal assignment set offers a higher resolution in obtaining the \textit{p}-value.
To achieve this, \cite{puelz2022graph} proposed the concept of \emph{null exposure graph} (NEG), which is a bipartite graph between units and assignments with an edge between unit $i\in\bbU$ and assignment $\bz\in\cZ$ if and only if unit $i$ is exposed to either $a$ or $b$ under $\bz$.
Encoding the impact of treatment assignments on units in terms of the observability of $\theta_i$'s under a given treatment assignment, NEG offers a way to better select focal units and focal assignments.
For example, every biclique in NEG defines a set of appropriate focal units and focal assignments.
Such an insight leads to a series of CRTs based on biclique decomposition of NEG, whose efficiency is further improved by \cite{yanagi2022improving} via power analysis.
More recently, \cite{zhong2024unconditional} expanded the core concept of CRTs into a more general framework named \textit{partial null randomization tests} (PNRTs), which reports an ensemble \textit{p}-value by averaging multiple \textit{p}-values obtained from a series of CRTs. 
Each CRT utilizes a small focal assignment set comprising only two focal assignments, yet incorporates an enlarged set of focal units including all focal units compatible with the specific pair of focal assignments.
In PNRTs, the larger focal unit set of each individual CRT enhances testing power, while the ensemble \textit{p}-value, obtained through averaging, addresses the issue of extremely low resolution in the individual \textit{p}-values.

Despite these efforts, randomization tests based on the idea of conditional inference still suffer from diminishing power.
For instance, consider a scenario where a high proportion of treatment $z_i = 1$ arises from ethical concerns (e.g., the treatment significantly benefits participants). This typically leads to a high proportion of exposure $m_i(\bz) = 2$. If the goal is to test for no spillover effect $H_0^{{0,1}}$, the proportion of units with observable potential outcomes under this null hypothesis, $N_{0,1}^\obs / N$, tends to be small for many assignments $\bz \in \mathcal{Z}$. This results in a sparse NEG, posing significant challenges in identifying sufficiently large bicliques within the NEG to conduct a randomization test with realistic power.

\subsection{Imputation-based \textit{p}-value for Testing a Complex Null Hypothesis}
For a parametric statistical model $f_\btheta(x)$ with $\btheta\in\Theta$ as the vector of parameters, consider testing the following complex null hypothesis
$H_0: \btheta \in \Theta_0,$
where $\Theta_0$ is a non-empty subset of the parameter space $\Theta$.
Given a collection of independent and identically distributed (i.i.d.) samples $\bX^{\obs}=(X_1,\cdots,X_n)$ from model $f_\btheta(x)$, researchers often test $H_0$ based on a parameter-dependent test statistics $D(\bX^{\obs},\btheta)$. 
In case that $\Theta_0$ contains only one element and thus $\btheta$ is fully specified under $H_0$, the \textit{p}-value for testing $H_0$ can be naturally established as below:
\begin{equation}\label{eq:classicp}
p(\bX^{\obs};\btheta) = \bbP_\bX\left(D(\bX,\btheta) \geq D(\bX^{\obs},\btheta)\right),
\end{equation}
where the probability $\bbP_{\bX}$ is derived by random sample 
$\bX\sim f_\btheta(x)$.
In case that $\Theta_0$ contains more than one element and thus $\btheta$ is not fully specified under $H_0$, the most classic way to establish the \textit{p}-value for $H_0$ is to find a proper estimate of $\btheta$ (e.g., the MLE) first, and then replace the unspecified $\btheta$ in \eqref{eq:classicp} by its estimate $\hat\btheta$, resulting in the classic plug-in \textit{p}-value (PIP) $p(\bX^{\obs};\hat\btheta)$.

From the Bayesian perspective, $p(\bX^{\obs};\hat\btheta)$ imputes the unspecified parameter $\btheta$ with a point estimation based on the observed data, which is obviously suboptimal.
A natural alternative is to impute $\btheta$ based on its posterior distribution under $H_0$.
Based on this insight, \cite{meng1994posterior} and \cite{gelman1996posterior} proposed to test $H_0$ by the \emph{posterior predictive \textit{p}-value} (PPP) defined as: 
\begin{equation}\label{eq:ppp}
{p}_B(\bX^{\obs};\Theta_0)=\bbE\left[p(\bX^{\obs};\btheta)\mid \bX^{\obs};H_0\right]=\int p(\bX^{\obs};\btheta)\pi(\btheta\mid\bX^{\obs},\Theta_0)d\btheta,
\end{equation}
where $\pi(\btheta\mid\bX^{\obs},\Theta_0)\propto f(\bX^{\obs}\mid\btheta)\pi(\btheta\mid \Theta_0)$ is the posterior distribution of $\btheta$ given $\bX^{obs}$ and a prior distribution $\pi(\btheta\mid\Theta_0)$ defined over $\Theta_0$. 
Notably, because typically $\pi(\btheta\mid\bX^{\obs}, \Theta_0)$ and $\hat{\btheta}$ both converge to the true parameter $\btheta$ as the sample size approaches infinity, the PPP ${p}_B(\bX^{\obs};\Theta_0)$ and the PIP ${p}(\bX^{\obs};\hat\theta)$ often shares the same asymptotical behavior.
A key advantage of PPP over PIP is its ability to address uncertainty in specifying $\btheta$ via the posterior distribution $\pi(\btheta \mid \bX^{\obs}, \Theta_0)$, which makes the inference procedure more robust. 

Many causal inference problems with complicated structures conform to the framework of PPP.
For example, \cite{rubin1998more} utilized PPP to improve the power of FRT in randomized experiments with one-sided noncompliance. 
Under the null hypothesis of no treatment effect for compliers, the unknown compliance types of units serve as the nuisance parameter $\btheta$, whose value can be imputed based on its posterior predictive distribution under a parametric model with an appropriate prior distribution. 
Averaging the pristine \textit{p}-values across the imputed datasets according to Eq. \eqref{eq:ppp}, 
PPP is obtained to summarize the observed evidence against the null hypothesis. 
Earlier efforts of utilizing multiple imputation based on posterior predictive distribution for Bayesian model checking were summarized in \cite{guttman1967use} and \cite{rubin1981estimation,rubin1984bayesianly}.
In this paper, we aim to utilize the imputation idea to address the unobservable potential outcomes in randomization tests with interference.

\section{Imputation-based Randomization Test 
}\label{sec:PPP}

\subsection{Methodology}

Following the spirit of PPP, we propose the following imputation-based randomization test.
First, we impute the missing elements of the partially observed $\btheta$ according to an imputation distribution $\pi$ to get a complete version of $\btheta$ referred to as $\btheta^{\imp}$ hereafter, and conduct the classic FRT based on the single imputation for the \textit{p}-value below:
\begin{equation}\label{eq:FRT-pValue-thetaImp}
p_{\SI}(\bZ^{\star};\btheta^{\imp}) 
= \bbP_\bZ\left(T(\bZ,\btheta^{\imp}) \geq T(\bZ^{\star},\btheta^{\imp})\right),
\end{equation} 
with the partially observed $\btheta$ in Eq. \eqref{eq:FRT_pValue} substituted by the fully imputed $\btheta^{\imp}$.
Next, we average $p_{\SI}$ according to the imputation distribution $\pi(\cdot \mid\btheta^\obs)$, which is dependent on the realized assignment and observed outcomes, to get the following ensemble \textit{p}-value: 
\begin{align}\label{eq:RPPP-Interference}
p_{\MI}(\bZ^{\star};\btheta^\obs)
&= 
\bbE\left(p_{\SI}(\bZ^{\star};\btheta^{\imp})\right)=\int p_{\SI}(\bZ^{\star};\btheta^{\imp})\pi(\btheta^{\imp}\mid\btheta^\obs)d\btheta^{\imp},
\end{align}
and reject $H_0^{\{a,b\}}$ if $p_{\MI}$ is smaller than a pre-specified significance level $\alpha$.
We term the above testing procedure as \textit{imputation-based randomization test} (IRT).
Considering that it's difficult to calculate $p_{\MI}(\bZ^{\star};\btheta^\obs)$ exactly because the involved FRT \textit{p}-values typically have no analytical form, we can approximate $p_{\MI}(\bZ^{\star};\btheta^\obs)$ via Monte-Carlo simulation, as suggested by \cite{rubin1998more}, after conducting the randomized experiment according to the realized assignment $\bZ^{\obs}$.
The algorithm below summarizes the procedure of IRT. Define $\bI(\cdot)$ as the indicator function.
\begin{algorithm}
\KwIn{Observed assignment $\bZ^\obs$, observed elements of nuisance parameters $\btheta^\obs$, imputation distribution $\pi(\cdot \mid\btheta^\obs)$, treatment assignment mechanism $\bF_\cZ$, significance level $\alpha$, number of simulations $K$.}
\For{$k = 1$ \KwTo $K$}{
(Imputation) Sample $\btheta^{\imp}_{(k)}$ from the imputation distribution $\pi(\cdot \mid\btheta^\obs)$.
\\
(Randomization)
Randomly sample $\bz_{(k)}$ from $\bF_\cZ$.\\
Calculate $R_k = \bI\big(T(\bz_{(k)},\btheta^{\imp}_{(k)}) \geq T(\bZ^{\star},\btheta^{\imp}_{(k)})\big)$.\\
}

(Averaging) $\hat{p}_{\MI}(\bZ^{\star};\btheta^\obs)=
K^{-1}\sum_{k=1}^K R_k$.\\
\KwOut{\textit{p}-value $\hat{p}_{\MI}(\bZ^{\star};\btheta^\obs)$; Reject $H_0^{\{a,b\}}$ if $\hat{p}_{\MI}(\bZ^{\star};\btheta^\obs) \leq \alpha$.}
\caption{Imputation-based Randomization Test (IRT)}
\end{algorithm}

Apparently, the imputation distribution $\pi$ plays an important role in the construction of IRT.
Assuming that units in $\bbU$ are random samples from a super population $\cP$ with $g(\cdot)$ as the marginal density of $\theta_i$ for $i\in\bbU$, the structure of the hypothesis testing problem defined in Eq. \eqref{eq:contrast} naturally suggests the following strategy to impute $\btheta=(\theta_1,\cdots,\theta_N)$, given its observed version $\btheta^\obs$:
we set $\theta_i^{\imp}=\theta_i^{\obs}$ for $i\in\bbU^{\star}_{a,b}$, and we impute $\theta_i^{\imp}$ with a random sample from $g(\cdot)$ for $i\notin\bbU^{\star}_{a,b}$, leading to the oracle imputation distribution
\begin{equation}\label{eq:true}
{ \pi_o(\btheta^{\imp} \mid \btheta^\obs)}
= \prod_{i\notin\bbU^{\star}_{a,b}} g(\theta_i^{\imp})\cdot 
{\prod_{i\in\bbU^{\star}_{a,b}} \delta(\theta_i^{\imp} - {\theta_i^\obs})},
\end{equation}
where $\delta(\cdot)$ is the Dirac delta function.
We refer to the IRT based on $\pi_o(\btheta^{\imp} \mid \btheta^\obs)$ as IRT$_\o$.

{In practice, however, it's often the case that distribution $g(\cdot)$ is unknown and needs to be estimated based on the observed subset in $\btheta^{\obs}$, namely $\Tilde{\btheta}^\obs=\{\theta_i^\obs\}_{i\in\bbU^\obs_{a,b}}$ according to Eq. \eqref{eq:thetaobs}.
Let $\hat{g}(\cdot\mid \Tilde\btheta^\obs)$ be the estimated marginal distribution of $\theta_i$, we get the following imputation distribution in action:
\begin{equation}\label{eq:estimate}
\hat{\pi}(\btheta^{\imp} \mid \btheta^\obs) = \prod_{i\notin\bbU^{\star}_{a,b}} \hat{g}(\theta_i^{\imp}\mid\Tilde\btheta^{\obs})\cdot \prod_{i\in\bbU^{\star}_{a,b}} 
\delta(\theta^{\imp}_{i} - \theta_i^\obs).
\end{equation}
When distribution $g(\cdot)$ has a parametric form with $\bgamma$ as the model parameters, we can learn $g(\cdot)$ from $\Tilde\btheta^{\obs}$ by inferring its parameter $\bgamma$ in either frequentist or Bayesian manner.
We term this version of IRT as IRT$_\p$.

If no appropriate parametric model is available, we can estimate $g(\cdot)$ via non-parametric approaches instead.
For example, we can approximate the unknown $g(\cdot)$ by the empirical distribution of the observed potential outcomes $\Tilde\btheta^{\obs}$, i.e.,
\begin{equation}\label{eq:randomsample2}
\hat{g}_e(\theta\mid \Tilde\btheta^\obs) 
= \frac{1}{N^{\star}_{a,b}} \sum_{i\in\bbU_{a,b}^\obs}\delta (\theta - \theta_i^\obs).
\end{equation}
Alternatively, when $g(\cdot)$ is known to be a continuous distribution, we can rely on the smoothed version of  the empirical distribution $\hat g$ instead as suggested by \cite{van1994weak}, i.e.,
\begin{equation}\label{eq:kernel_function}
\hat{g}_s(\theta\mid \Tilde\btheta^\obs) = \frac{1}{N_{a,b}^\obs}\sum_{i\in\bbU_{a,b}^\obs}K\left(\frac{\theta-\theta_i^\obs}{h_N}\right),
\end{equation}
where $K(\cdot)$ is a symmetric kernel function with bandwidth $h_N$ satisfying $\int K(u)du=1$, $\int u^2 K(u) du < \infty$, and $K(u) = K(-u)$.
We refer to the IRTs based on $\hat{g}_e$ and $\hat{g}_s$ as IRT$_\e$ and as IRT$_\s$, respectively.

\subsection{Theoretical Results}

In this section, we present the theoretical results for IRT under the null hypothesis in Eq. \eqref{eq:contrast}, when the oracle imputation distribution is precisely known or needs to be estimated based on the observables.
Some of the results are closely related to the results for PPP in \citet{meng1994posterior}.
A critical difference between IRT and PPP, however, is that the imputation distribution in IRT would not converge to a point mass, as in the classic PPP, with the increase of sample size.
Such a fact leads to some unique issues in the theoretical analysis of IRT, making it non-trivial to establish the validity of IRT.
First, we start with the simple case where IRT is conducted with the oracle imputation distribution, i.e., IRT$_\o$.

\begin{theorem}
\label{thm:oracle}
Assume that 
(i) the null hypothesis in Eq. \eqref{eq:contrast} holds with $Y_i(a) = Y_i(b) =\theta_i$, 
(ii) $\theta_i$ are i.i.d. samples from distribution with probability density $g(\cdot)$,
(iii) $\bZ^\obs \sim \bF_\cZ$,
and 
(iv) the imputation of $\btheta^{\imp}$ is according to the oracle imputation distribution $\pi_o(\btheta^{\imp} \mid\btheta^\obs)$ as defined in Eq. \eqref{eq:true}.
Then $p_{\MI}(\bZ^{\star};\btheta^{\obs})$ is stochastically less variable than a uniform distribution but with the same mean, which means that:
(i) $\bbE_{(\bZ^\obs,\btheta)}[p_{\MI}(\bZ^{\star};\btheta^{\obs})] = \bbE_U[U] = 1/2$, and (ii) $\bbE_{(\bZ^\obs,\btheta)}\left[h\left(p_{\MI}(\bZ^{\star};\btheta^{\obs})\right)\right] \leq \bbE_U[h(U)]$ for all convex functions $h$ on $[0,1]$, 
where $U$ is a random sample from the uniform distribution on $[0,1]$, the expectation $\bbE_{(\bZ^\obs,\btheta)}$ is taken with respect to the joint distribution of $(\bZ^{\star},\btheta)$, and $\bbE_U$ is taken with respect to $U$. 
\end{theorem}

{Theorem \ref{thm:oracle} shows that IRT$_\o$ is more centered around $1/2$ than $U\sim \unif(0,1)$ when the null hypothesis holds. 
The proof is detailed in the Supplementary Material. 
The theorem immediately leads to the following corollary based on Lemma 1 in \citet{meng1994posterior}.
}

\begin{corollary}\label{cor:typeI}
{Under the same assumptions in Theorem \ref{thm:oracle}, we have}
$$
\bbP\left(p_{\MI}(\bZ^{\star};\btheta^{\obs}) \leq \alpha\right) \leq 2\alpha,\quad { \forall\ \alpha\in(0,1)},
$$
where $\bbP$ is the probability with respect to $\bZ^\obs \sim \bF_\cZ$ and $\theta_1,\ldots,\theta_N \stackrel{i.i.d.}\sim g(\cdot)$.
\end{corollary}

Corollary \ref{cor:typeI} ensures that IRT$_\o$ can control the type I error rate at most twice the significance level, i.e., $\alpha$, a result that is well-known for PPP.
Although such a result cannot be further improved to more positive forms from the theoretical perspective, e.g., $\bbP\left(p_{\MI}(\bZ^{\star};\btheta^{\obs}) \leq \alpha\right) \leq \alpha$ for all $\alpha>0$, according to the counterexample in \cite{dahl2006conservativeness}, our empirical studies in Sections \ref{sec:simulations} and \ref{sec:realdata} show that the type I error rate of IRT$_\o$ and its extensions are often well controlled at $\alpha$. 
Interestingly, such a phenomenon also occurs in other model-free, finite-sample exact inference methods, such as cross-conformal prediction, jackknife+, and CV+ \citep{vovk2018cross, vovk2020combining, barber2021predictive,guan2024conformal}. 
For instance, \citet{barber2021predictive} recommended the jackknife+ prediction interval for its typical empirical coverage of $1-\alpha$, although theoretically, it can only guarantee $1-2\alpha$ coverage due to certain pathological cases.
These practices provide extra support for the application of IRT.

Next, we focus on the more challenging cases where IRT is conducted with the estimated imputation distribution as shown in Eq. \eqref{eq:estimate}.
Define
\[\begin{split}
\Delta(\bZ^{\star};\btheta^{\obs})=\int p_{\SI}(\bZ^{\star};\btheta^{\imp})\left[\hat{\pi}(\btheta^{\imp}\mid \btheta^{\obs})-\pi_o(\btheta^{\imp}\mid \btheta^{\obs})\right] d \btheta^{\imp}
\end{split}\]
as the ensemble \textit{p}-value difference between the two versions of IRT with the estimated and oracle imputation distributions.
Intuitively, if the distribution of $\Delta(\bZ^\obs;\btheta^\obs)$, which is determined by the joint distribution of $(\bZ^\obs,\btheta)$, concentrates at 0, the IRT under the estimated imputation distribution will closely resemble IRT$_\o$ with similar properties. 
The following lemma provides a positive answer to the above intuition, ensuring that if $\Delta(\bZ^{\star};\btheta^{\obs})$ converges to 0 in probability, the asymptotic type I error rate of IRT with the corresponding estimated imputation distribution is also bounded by twice the significance level.
\begin{lemma}\label{lma:similar}
Assume that 
(i) the null hypothesis in Eq. \eqref{eq:contrast} holds with $Y_i(a) = Y_i(b) =\theta_i$, 
(ii) $\theta_i$ are i.i.d. samples from distribution with probability density $g(\cdot)$,
and (iii) $\bZ^\obs \sim \bF_\cZ$.
If $\Delta(\bZ^\obs;\btheta^\obs) = o_p(1)$, then for $p_{\MI}(\bZ^{\star};\btheta^{\obs})$ with imputation Eq. \eqref{eq:estimate}, we have:
\begin{equation}\label{eq:AsympError}
\limsup_{N\rightarrow\infty}\bbP\left(p_{\MI}(\bZ^{\star};\btheta^{\obs}) \leq \alpha\right) \leq 2\alpha,\quad { \forall\ \alpha\in(0,1)}.
\end{equation}
\end{lemma}

Denote the sampling distribution of $N_{a,b}^\obs$ derived by $\bZ^\obs \sim \bF_\cZ$ as $\bF_N$, and the corresponding probability mass function (p.m.f.) as
\begin{equation}\label{eq:pi_N}
\bbP_N(N_{a,b}^\obs = n)= \sum_{\bz: |\bbU_a(\bz)\cup\bbU_b(\bz)|=n} \bbP_{\bZ}(\bZ=\bz),
\end{equation}
where $\bbP_\bZ$ is defined over $\bZ\sim \bF_\cZ$. 
The following theorem provides sufficient conditions under which 
$\Delta(\bZ^{\star};\btheta^{\obs}) = o_p(1)$, and thus Eq. \eqref{eq:AsympError} holds.

\begin{theorem}\label{thm:SufficientCondition}
Under the same assumptions in Lemma \ref{lma:similar}, 
if $g$ and $\hat{g}$ correspond to continuous distributions, 
a sufficient condition for $\Delta(\bZ^{\star};\btheta^{\obs}) = o_p(1)$, and thus Eq. \eqref{eq:AsympError}, is 
\begin{equation}\label{eq:condition}
\bbE\left|\prod_{i=N_{a,b}^\obs+1}^{N}\frac{\hat{g}\left(\vt_i\mid \bmvt_{1:N_{a,b}^\obs}\right)}{g(\vt_i)}-1\right| = o(1)\text{ as }N \rightarrow \infty,
\end{equation}
where $\bbE$ is the expectation taken with respect to
$\vt_1,\ldots,\vt_N$ i.i.d. with probability density function $g(\cdot)$, $\bmvt_{1:N_{a,b}^\obs} = \{\vt_1,\ldots,\vt_{N_{a,b}^\obs}\}$ and $N_{a,b}^\obs\sim \bF_N$. 
If $g$ and $\hat{g}$ correspond to discrete distributions with corresponding p.m.f. \textit{p} and $\hat{p}$, the sufficient condition is 
\begin{equation}\label{eq:condition_discrete}
\bbE\left|\prod_{i=N_{a,b}^\obs+1}^{N}\frac{\hat{p}\left(\vt_i\mid \bmvt_{1:N_{a,b}^\obs}\right)}{p(\vt_i)}-1\right| = o(1)\text{ as }N \rightarrow \infty,
\end{equation}
where $\bbE$ is the expectation taken with respect to $\vt_1,\ldots,\vt_N$ i.i.d. with probability density function $g(\cdot)$, $\bmvt_{1:N_{a,b}^\obs} = \{\vt_1,\ldots,\vt_{N_{a,b}^\obs}\}$ and $N_{a,b}^\obs\sim \bF_N$.
\end{theorem}

The above sufficient conditions imply that when the likelihood ratio between the estimated distribution $\hat{g}$ and the true distribution $g$ on the unobservables converges to 1 in $L^1$, IRT with the estimated imputation distribution will exhibit similar performance as IRT$_\o$, whose behavior has been established previously.
The corollary below shows that in a simple case where $g(\cdot)$ represents a normal distribution with known variance, the sufficient conditions can be theoretically validated under specific conditions.

\begin{corollary}\label{cor:normal_example}
Under the same assumptions in Lemma \ref{lma:similar},
let $\theta_i$ be i.i.d. samples from $\cN(\mu,\sigma^2)$ with known variance $\sigma^2$ and unknown mean $\mu$.
Suppose the sampling distribution $\bF_N$ of $N_{a,b}^\obs$ is a point mass distribution, so that the missing rate of nuisance parameters, defined as $r_N^\mis = 1-N_{a,b}^\obs/N$, is a constant depending on $N$.
If we estimate the unknown density $g(\cdot)$ with its posterior predictive distribution $\hat{g}(\cdot\mid \Tilde\btheta^\obs)$, given the observed outcomes $\Tilde\btheta^\obs$ and a normal prior for $\mu$, 
then \eqref{eq:AsympError} holds when $r_N^\mis = o(1)$. 
\end{corollary}

For more complex cases, theoretical verification of the sufficient conditions in Theorem \ref{thm:SufficientCondition} becomes challenging, but can be approximated by numerical verification when $r_N^\mis$ converges to 0 at a sufficiently fast rate, e.g., $O(N^{-1/2})$.
Further details are provided in the Supplementary Material.

\section{Simulations}\label{sec:simulations}

\subsection{Clustered Interference}
\label{sec:sim1}

In this section, we examine the proposed method under various interference mechanisms via simulation.
First, we consider the clustered interference as defined in Example \ref{eg:clustered}.
For a group of $N = 300$ units equally divided into $K \in \{30, 75, 150\}$ clusters, we are interested in testing the null hypothesis of no spillover effect $H_0^{\{0,1\}}$.
Since the potential outcome $Y_i(2)$ is irrelevant to our interest, we only generate potential outcomes $Y_i(0)$ and $Y_i(1)$ by
\begin{equation}\label{eq:generate_PO}
Y_i(0)\sim\mathcal{N}(0,1),\ Y_i(1) = Y_i(0) + \tau,\ i=1,\ldots,N, 
\end{equation}
where $\tau \in \{0,0.1,\cdots,1\}$.
To identify the spillover effect under clustered interference, we consider the two-stage randomized experiment suggested by \cite{hudgens2008toward}, where we randomly assign $\lfloor K/2 \rfloor$ clusters to the treatment group and the remaining clusters to the control group in the first stage, and then randomly select one unit from each cluster in the treatment group to receive the treatment in the second stage with all the other units untreated. 
For each of the $3\times 11$ configurations of $(K,\tau)$, we generate 10 independent datasets according to the data simulation model in Eq. \eqref{eq:generate_PO}.
For each simulated dataset with a specific set of potential outcomes, we conduct 1,000 parallel randomized experiments with the treatment assignment $\bZ$ randomly specified based on the randomized design $\bF_\cZ$ according to the two-stage experiment.

Using the difference-in-means in Eq. \eqref{eq:CRT_TestStat} as the test statistic with the significance level to reject the null hypothesis set to $\alpha = 0.05$, we compare the proposed IRT methods, including IRT$_\o$, IRT$_\p$, IRT$_\e$ and IRT$_\s$, to a group of CRT methods with focal units and assignments selected by various algorithms (including CRT$_\o$, CRT$_\b$, CRT$_{\db}$ and CRT$_{\pb}$) and two PNRT methods (namely PNRT$_\p$ and PNRT$_\m$).
Detailed configurations of the 10 involved competing methods can be found in Table \ref{tab:methods}.
For each of these competing randomization tests, we measure its type I error rate via simulated datasets with $\tau=0$, and power via simulated datasets with $\tau\neq 0$.
\begin{table}[t]
    \centering
    \begin{tabular}{cccc}
\hline
 \multirow{2}{*}{Methods}       &  \multirow{2}{*}{Description}           & \multicolumn{2}{c}{Interference Type}\\ \cline{3-4}
 &  & Clustered & Spatial \\
\hline
\rowcolor{gray!20}
CRT$_\o$ & \parbox{10.2cm}{CRT with \textbf{o}ne focal unit per cluster \citep{basse2019randomization}} & \ding{51} &  \ding{55}\\[0.5em]
CRT$_\b$ & \parbox{10.2cm}{\vspace{0.2em} CRT with general \textbf{b}iclique decomposition \citep{puelz2022graph}}& \ding{51} & \ding{51} \\
\rowcolor{gray!20}
CRT$_{\db}$ & \parbox{10.2cm}{\vspace{0.2em} CRT with \textbf{d}esign-assisted \textbf{b}iclique decomposition \citep{puelz2022graph}} & \ding{51} & \ding{55} \\
CRT$_{\pb}$ & \parbox{10.2cm}{\vspace{0.2em} CRT with \textbf{p}ower-analysis-assisted \textbf{b}iclique decomposition \citep{yanagi2022improving}}& \ding{51}  &  \ding{51}\\
\rowcolor{gray!20}
PNRT$_\p$ & \parbox{10.2cm}{\vspace{0.2em} The \textbf{p}airwise comparison-based PNRT \citep{zhong2024unconditional}}& \ding{51}  & \ding{51}\\
PNRT$_\m$ & \parbox{10.2cm}{The \textbf{m}inimization-based PNRT \citep{zhong2024unconditional}}  & \ding{51}& \ding{51}\\
\rowcolor{gray!20}
IRT$_\o$ & \parbox{10.2cm}{IRT imputed with the \textbf{o}racle distribution} & \ding{51} & \ding{51}\\ 
\vspace{0.2em} IRT$_\p$ & \parbox{10.2cm}{\vspace{0.2em} IRT imputed with a \textbf{p}arametric model} & \ding{51} &  \ding{51}\\ 
\rowcolor{gray!20}
IRT$_\e$ & \parbox{10.2cm}{IRT imputed with the \textbf{e}mpirical distribution} & \ding{51} & \ding{51}\\ 
IRT$_\s$ & \parbox{10.2cm}{IRT imputed with a \textbf{s}mooth kernel density function}& \ding{51}  &  \ding{51}\\ 
\hline
\end{tabular}
\caption{Methods evaluated in the simulation study.}
\label{tab:methods}
\end{table}
 
We present the type I error rate across 10 datasets and the average power of the 10 competing methods in Fig. \ref{fig:cluster_normal}.
For cases where $K=150$, however, the results of CRT$_\b$ and CRT$_{\pb}$ are absent because these two methods become infeasible in these cases. 
When $K=150$, the average missing rate of $\theta_i$ reaches up to 25\%, leading to a null exposure graph with a density lower than 0.75. 
Consequently, it is impractical to find a biclique decomposition that meets the minimum size requirements established by \cite{puelz2022graph} and \cite{yanagi2022improving}, which specify at least 20 focal assignments and 20 focal units, rendering CRT$_\b$ and CRT$_{\pb}$ infeasible.
From the results, we make the following observations.
First, while IRT and CRT methods successfully control the type I error rate around 0.05 across all 10 datasets, the two PNRT methods fail to do so and tend to yield type I error rates that are much smaller than 0.05 in all cases.
Although our theory guarantees only a $2\alpha$ bound of the type I error rate for IRT methods asymptotically, these simulation results suggest that IRT can often successfully control the type I error rate below $\alpha$ in practice.
Second, the IRT methods demonstrate significantly higher power than CRT and PNRT methods across all cases, especially for cases with a large number of clusters $K$.
Although CRT$_\o$ and CRT$_{\db}$ also yield competitively high power in these cases, they are specifically designed for clustered interference only and can not be extended to other settings.
Third, the IRT methods based on estimated distribution -- namely IRT$_\e$, IRT$_\p$ and IRT$_\s$ -- exhibit performance similar to IRT$_\o$, even though the missing rate of nuisance parameters is far from 0.
These results confirm that the proposed IRTs are powerful tools for casual inference in presence of cluster interference.

Additional experiments where $Y_i(0)$ values are generated from heavier tail distributions, such as $\chi^2$ or $t$ distributions, but still modeled as a normal distribution, are reported in the Supplementary Materials, showing that IRT$_\p$ is robust to model mis-specification.

\begin{figure}[t]
\centering
\includegraphics[width=0.8\linewidth]{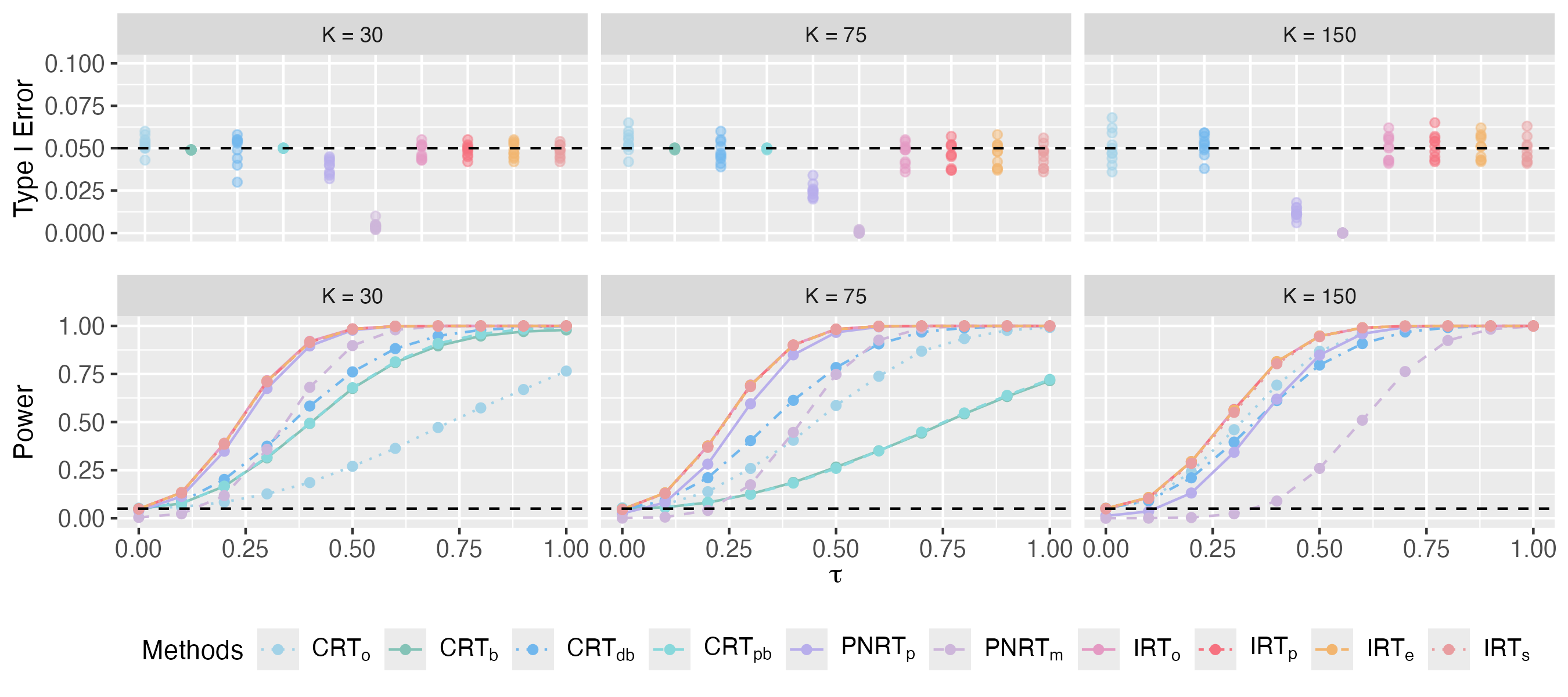}
\caption{Simulation results under clustered interference. The type I error rate across 10 datasets and the average power over these datasets are visualized for the 10 competing methods under different specifications of causal effect $\tau$ and number of clusters $K$. The horizontal dashed line indicates the significance level of $\alpha = 0.05$.}
\label{fig:cluster_normal}
\end{figure}

\subsection{Spatial Interference}

Next, we consider the spatial interference as defined in Example \ref{eg:spatial}.
For a group of $N=1,000$ units with spatial interference, the potential outcomes are simulated according to Eq. \eqref{eq:generate_PO}.
We adopt the similar setting as in \cite{yanagi2022improving} to specify the interference mechanism: each unit $i$ is associated with a 2-dimensional coordinate $\x_i\in\bbR^2$ according to
\[\begin{split}
\x_i \sim& \mathcal{N}\big((0.5,0.5)^\T,0.1^2 I_2\big)\cdot I(1\leq  i\leq500)+\cN\big((0.25,0.75)^\T,0.075^2 I_2\big)\cdot I(500<i\leq800)\\
&+ \cN\big((0.3,0.3)^\T,0.075^2 I_2\big)\cdot I(800<i\leq 1000),
\end{split}\]
where $I_2$ denotes the $2\times2$ identity matrix,
and allow interference to occur between two unit $i$ and $j$ if and only if their Euclidean distance $d(i,j)=||\x_i-\x_j||_2\leq r$, with the distance of interference $r \in \{0.005,0.01,0.02\}$.
For each configuration of $(r,p,\tau)$, we generate 10 independent datasets of sample size $N$ according to Eq. \eqref{eq:generate_PO}.
For each simulated dataset, the treatment assignment $\bZ=(Z_1,\ldots,Z_n)$ is sampled randomly for 1,000 times according to an independent Bernoulli trial where $Z_i\sim \text{Bernoulli}(p)$ for  $i=1,\ldots,N$ with $p\in\{0.2,0.5,0.8\}$.

\begin{figure}[!t]
\centering
\includegraphics[width=0.8\linewidth]{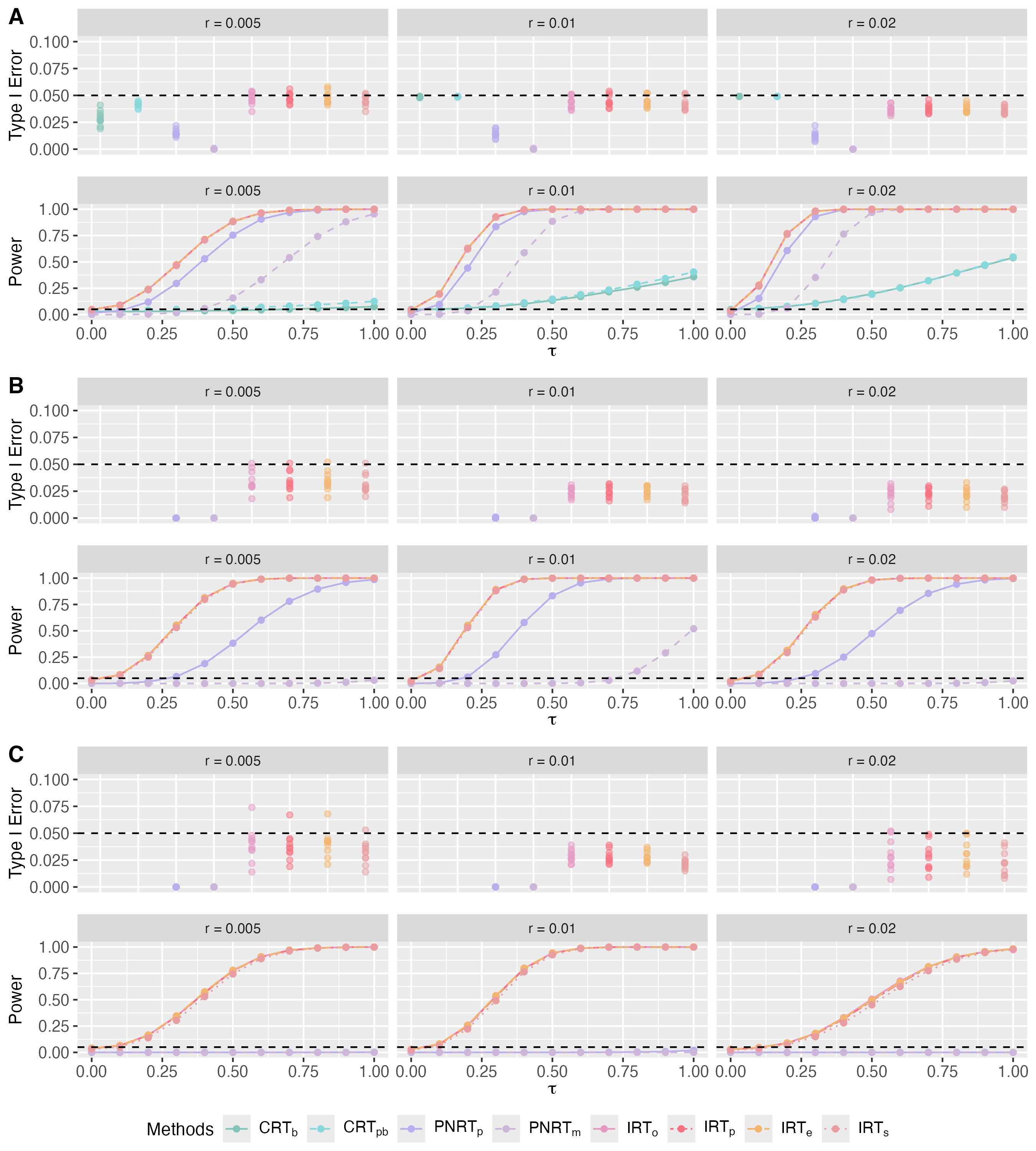}
\caption{Simulation results under spatial interference: A, B and C show the type I error rate and average power of the 10 competing methods for $p=0.2, 0.5$, $0.8$, under different specifications of casual effect $\tau$ and distance of interference $r$. The horizontal dashed line indicates the significance level of $\alpha = 0.05$.}
\label{fig:spatial_normal}
\end{figure}

Using the difference-in-means in Eq. \eqref{eq:CRT_TestStat} as the test statistic, 
we aim to test the null hypothesis of no spillover effect, i.e., $H_0^{\{0,1\}}$, with a significance level of $\alpha = 0.05$.
The type I error rate and average power across 10 datasets for the methods described in Section~\ref{sec:sim1} are 
measured for comparison.
The results are presented in Fig. \ref{fig:spatial_normal}, where CRT$_\o$ and CRT$_{\db}$ are not included due to their inapplicability under spatial interference.
In this figure, subfigures A, B, and C depict the type I error rate and power for $p=0.2$, $0.5$, and $0.8$, respectively. 
For $p=0.5$ and $0.8$, the results of CRT$_\b$ and CRT$_{\pb}$ are absent due to the high average missing rate of nuisance parameters, which can reach up to 50\% and 80\%.

These results show that the proposed IRTs outperform existing methods significantly under spatial interference as well as under clustered interference.
First, the proposed IRT methods effectively control the type I error rate and exhibit significantly higher power compared to CRTs and PNRTs, especially in the case of $p = 0.8$, where CRTs are not applicable and PNRTs have almost no power at all.
Second, similar to the results under clustered interference, the other IRT methods exhibit similar performance to IRT$_\o$ even with a large missing rate of nuisance parameters. 
Additional results for $Y_i(0)$ generated from $\chi^2(4)$ and  $t(4)$ are detailed in the Supplementary Material, which show that IRT methods remain robust across different distributions in the case of spatial interference as well.

Because all simulation results show that IRTs based on different estimated imputation distributions perform similarly to IRT$_\o$, we recommend using IRT$_\e$ in practice for its ease of implementation and independence from additional assumptions for the marginal distribution of $\theta_i$.

\section{Real Data Application}\label{sec:realdata}

\subsection{The Insurance Purchase Data}

We re-analyze the insurance purchase data collected from a two-round randomized experiment with multiple treatments conducted by \cite{cai2015social} to study the influence of social network among farmers on their decisions to purchase a new weather insurance product. 
In this experiment, $N=4,902$ households of rice farmers were randomly assigned to the first or the second rounds of insurance introduction sessions.
The households assigned to both rounds were randomized to receive either a simple or an intensive introduction to the insurance product.
The households assigned to the second round, however, were further  divided into three random groups and informed additional information about the purchase decisions of the first-round in three different ways: no additional information at all (NoInfo), summary information in terms of the overall purchase rate (Overall), or detailed information on concrete purchase decisions of all households in the first round (Indiv).
At the end, all involved households were randomized into $2+2\times 3=8$ treatment groups, as summarized in Fig \ref{fig:treatment}.
A stratified randomized experiment was conducted to randomly assign the 8 treatments to households.

\begin{figure}[t]
    \centering
    \includegraphics[width=0.8\linewidth]{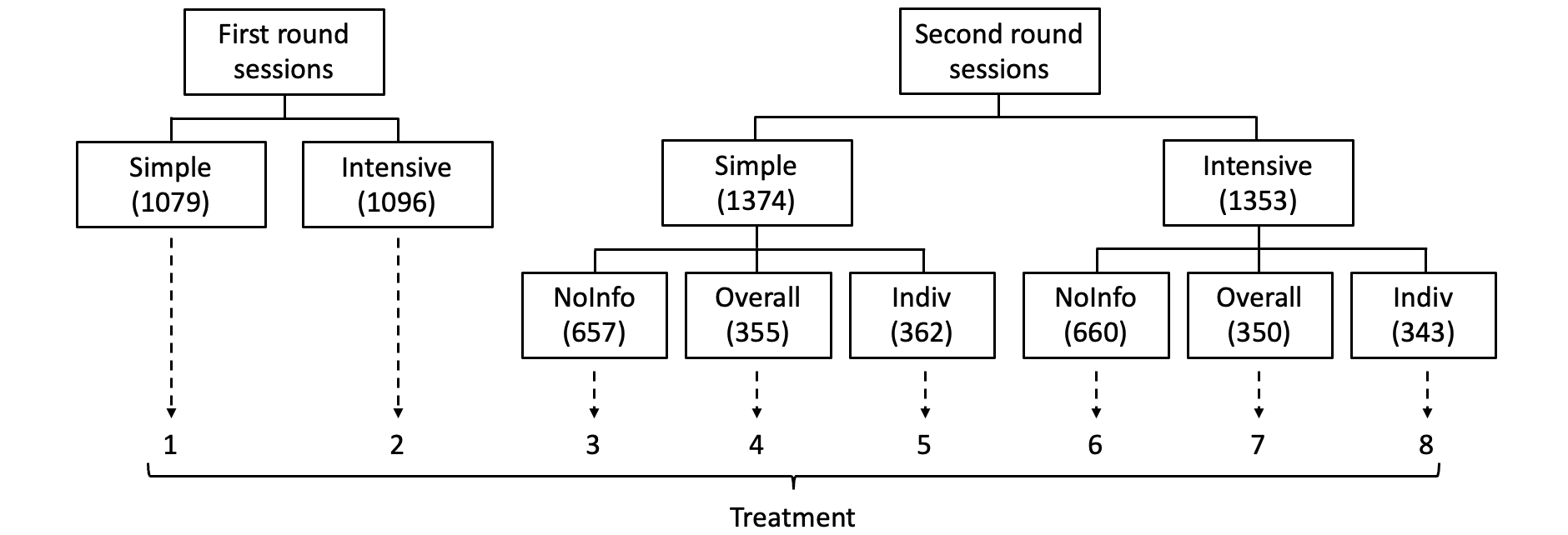}
    \caption{Eight treatment groups in the two-round experiment conducted by \cite{cai2015social}. Numbers in the brackets highlight how many households are assigned into these treatment groups in the experiment.}
\label{fig:treatment}
\end{figure}

Households were asked to make a purchase decision immediately after the insurance introduction sessions, making sure that there was \emph{no interference} between households in the same round.
However, there was a 3-day gap between the two rounds to allow households from the first round to personally discuss the insurance with their friends in the second round, implying a \emph{one-way interference} from the first round to the second round.
The social network of the 4,902 households was known in the experiment.
Let $\bW=(W_{ij})_{N \times N}$ be the adjacent matrix of the social network. 
We have $W_{ij}=1$ if two households $i$ and $j$ are friends, and  $W_{ij}=0$ otherwise. 
The purchase decisions of all households, i.e., $Y_i$'s, were recorded as the output of the experiment, where $Y_i=1$ if household $i$ made purchase and $Y_i=0$ otherwise.

\subsection{Testing the Spillover Effect by IRTs}

Containing rich ingredients, the insurance purchase data can be analyzed in different ways to answer different scientific questions.
Here, we are interested in testing the spillover effect from the first-round households to the second-round households.
To be concrete, let $\bbU=\{1,\dots,N\}$ be the $N=4,902$ households of interest, $\cT=\{1,\ldots,8\}$ be the treatment set showed in Fig.~\ref{fig:treatment}, $Z^\obs_i\in\cT$ be the observed treatment assignment of household $i\in\bbU$ and $\bZ^\obs=(Z^\obs_1,\ldots,Z^\obs_N)$.
Due to the possible interference between households through their social network, a household assigned to treatment $t\in\cT$ may receive distinct exposures depending on how much prior information its friend households can bring to it before treatment:
if some of its friend households get access to the intensive introduction of the insurance product a priori {in the first stage}, treatment $t$ would be conducted with strong prior information, leading to exposure $t^{(s)}$; 
if some of its friend households get access to the simple introduction a priori only, treatment $t$ would be conducted with weak prior information, leading to exposure $t^{(w)}$;
if none of its friend households get access to the simple or intensive introduction a priori, treatment $t$ would be conducted with no prior information at all, leading to exposure $t^{(n)}$.
Although we allow insurance content to spill over through social networks, for simplicity, we assume that households cannot obtain information about the first-round purchase decisions from social networks; they can only obtain this information from the experimenter as intended.
Based on these insights, we work on the following exposure mapping from $\cZ$ to exposure set $\cE=\cup_{t\in\cT}\{t^{(n)},t^{(w)},t^{(s)}\}$:
\begin{equation}\label{eq:RealData_ExposureMapping}
m_i(\bz) = z_i^{(a_i)},
\ \forall\ i\in\bbU,\ \bz\in\cZ,
\end{equation}
{where indicator $a_i$ takes value in set $\{n,w,s\}$ highlighting the level of prior information household $i$ receives before treatment}.
Note that because $a_i$'s are fully determined by the treatment assignment $\bz$ and the social network of the households, $m_i(\bz)$ in Eq.~\eqref{eq:RealData_ExposureMapping} is a well defined exposure mapping.

For treatment $t\in\cT$, we consider testing the following null hypothesis of no spillover effect, which is the contrast between $t^{(w)}$ and $t^{(s)}$ within treatment group $t$:
\begin{equation}\label{eq:hypothesisReal}
H_0^t: Y_i(t^{(w)}) = Y_i(t^{(s)}),\ 1\leq i\leq N.
\end{equation}
For this purpose, we use the difference-in-means between the $t^{(s)}$ exposure group and the $t^{(w)}$ exposure group as 
the estimator of the spillover effect and 
the test statistic for testing $H_0^t$ for treatment $t$, and establish two-sided \textit{p}-value of the test via various randomization tests.
For IRTs, we consider IRT$_\e$ and IRT$_\p$ with the following Beta-Binomial model for $\theta_i$'s:
$\theta_i \sim \text{Binomial}(1, q)$, $q \sim \text{Beta}(1, 1).$
For PNRTs, we try both PNRT$_\p$ and PNRT$_\m$ under the default setting.
For CRTs, however, we find that all members in this method family are not applicable to this real data example:
CRT$_\o$ and CRT$_\db$ cannot be applied to two-round experiments with multiple treatments and one-way interference, and CRT$_\b$ and CRT$_\pb$ are infeasible because the low density (e.g., 0.107 when $t=3$) of the null exposure graph makes the biclique decomposition infeasible.
Therefore, we only compare the performance of IRT$_e$ and IRT$_p$ versus PNRT$_\p$ and PNRT$_\m$ here.

\begin{table}[t]
\centering
\begin{tabular}{ccccccc}
\hline
\multirow{2}{*}{Treatment} & \multirow{1}{*}{Estimated} & \multicolumn{5}{c}{$\quad$\textit{p}-values reported by different randomization tests$\quad$}\\ \cline{3-7}
& spillover effect & CRT & PNRT$_\p$ &  PNRT$_\m$ & IRT$_\e$ & IRT$_\p$\\ \hline
1 & -0.009  & $-$ & 0.560 & 1.000 & 0.805 & 0.796\\
2 & -0.019 &  $-$ & 0.470 & 1.000 & 0.600 & 0.593\\ 
\cdashline{1-7}
3 & 0.133 &  $-$ & 0.423 & 1.000 & \textbf{0.002} & \textbf{0.006}\\
4 & -0.022 &  $-$ & 0.590 & 1.000 & 0.739 & 0.749\\
5 & 0.113 &  $-$ & 0.503 & 1.000 & 0.085 & 0.080\\ 
6 & 0.011 &  $-$ & 0.564 & 1.000 & 0.795 & 0.829\\
7 & 0.138 &  $-$ & 0.504 & 1.000 &  0.038 & 0.035\\
8 & 0.128 &  $-$ & 0.510 & 1.000 & 0.048 & 0.059\\
\hline
\end{tabular}
\caption{Testing results for the two-round experiment. The significance level is $\alpha = 0.05$, and the Bonferroni adjusted significance level is $\alpha / 8 = 0.00625$.}
\label{tab:results}
\end{table}

Table \ref{tab:results} summarizes the spillover effect for $t\in\{1,\cdots,8\}$ and the corresponding \textit{p}-values obtained by different methods.
From the table, we can see the following facts.
First, IRTs, as well as PNRTs, suggest that spillover effects are not significant for treatment $1$ and $2$, which is consistent with the domain knowledge of no interference within the same round.
Second, both IRT methods reveal significant spillover effects for treatment 3 and treatment 7, although only the spillover effect for treatment 3 is significant after considering the Bonferroni adjustment for multiple testing.
Since households in the treatment 3 group received only a simple introduction with no additional information on the purchase decision in the first round, it’s reasonable that leaked information from the social network would significantly influence their purchase decision.
However, both PNRT methods miss these signals, with all \textit{p}-values from PNRT$_\m$ far from 0.05.

\subsection{Simulation Based on Real Data}

To further assess the performance of the proposed methods in practice, we train a predictive regression model for the potential outcomes based on the real data and generate simulated data from the fitted model for performance evaluation and comparison.
To achieve this goal, we select 9 relevant covariates of the households according to \cite{cai2015social} as predictors to build a logistic regression model for the binary potential outcomes, resulting in a $4902 \times 9$ covariate matrix $\bX$. 
Because 6 of the 9 columns in $\bX$ have missing values, we would like to impute these missing values for convenient model training and data simulation.
Here, we follow the strategy suggested by \cite{zhao2024adjust} to impute all missing values in $\bX$ with 0, and include the binary missing data indicators (1 for missing elements and 0 otherwise) of the 6 incomplete covariates as additional covariates, resulting in an extra $4902 \times 6$ covariate matrix $\bM$. 
Moreover, dummy variables recording actual exposures received by households in the experiment, i.e., $\bE=(E_{i,e})_{i\in\bbU, e\in \cE}$, are also included in the model as covariates, where $E_{i,e}=1$ if household $i$ received exposure $e$ in the experiment.
With these covariates, we fit the logistic model for the observed outcomes $\bY^{\obs}$ as our base model for data generation: $\text{logistic}(\bY^{\obs} \sim 1+\bE+\bX+\bM).$
Simulation datasets can be generated based on the above base model to evaluate the performance of various methods for testing spillover effects.
For illustrative purpose, we focus on testing $H_0^t$ with $t=3$, which examines the spillover effect between exposures $3^{(s)}$ and $3^{(w)}$.
To be specific, let coef$(s)$ and coef$(w)$ be the coefficients of dummy variables $E_{:,3^{(s)}}$ and $E_{:,3^{(w)}}$ obtained in the fitted base model, and $\Delta(s,w)$ be their difference.
We modify the base model by resetting 
coef$(s)=$ coef$(w) + \Delta_{s,w} \cdot r$ for some $r \in (0,1)$,
and generate potential outcomes $\bY(3^{(s)})$ and $\bY(3^{(w)})$ of the same sample size as the original experiment according to the modified model.
If $r=0$, coef$(s)=$ coef$(w)$ in the modified model, and $H_0$ holds; if $r>0$, $H_0$ fails. 
Here, we adjust $r$ to align the differences in means of the generated $\bY(3^{(s)})$ and $\bY(3^{(w)})$ (denoted as $\tau$) to eight different values, ranging from 0 to 0.27, to simulate scenarios with varying signal strengths.

\begin{figure}[t]
\centering
\includegraphics[width=0.8\linewidth]{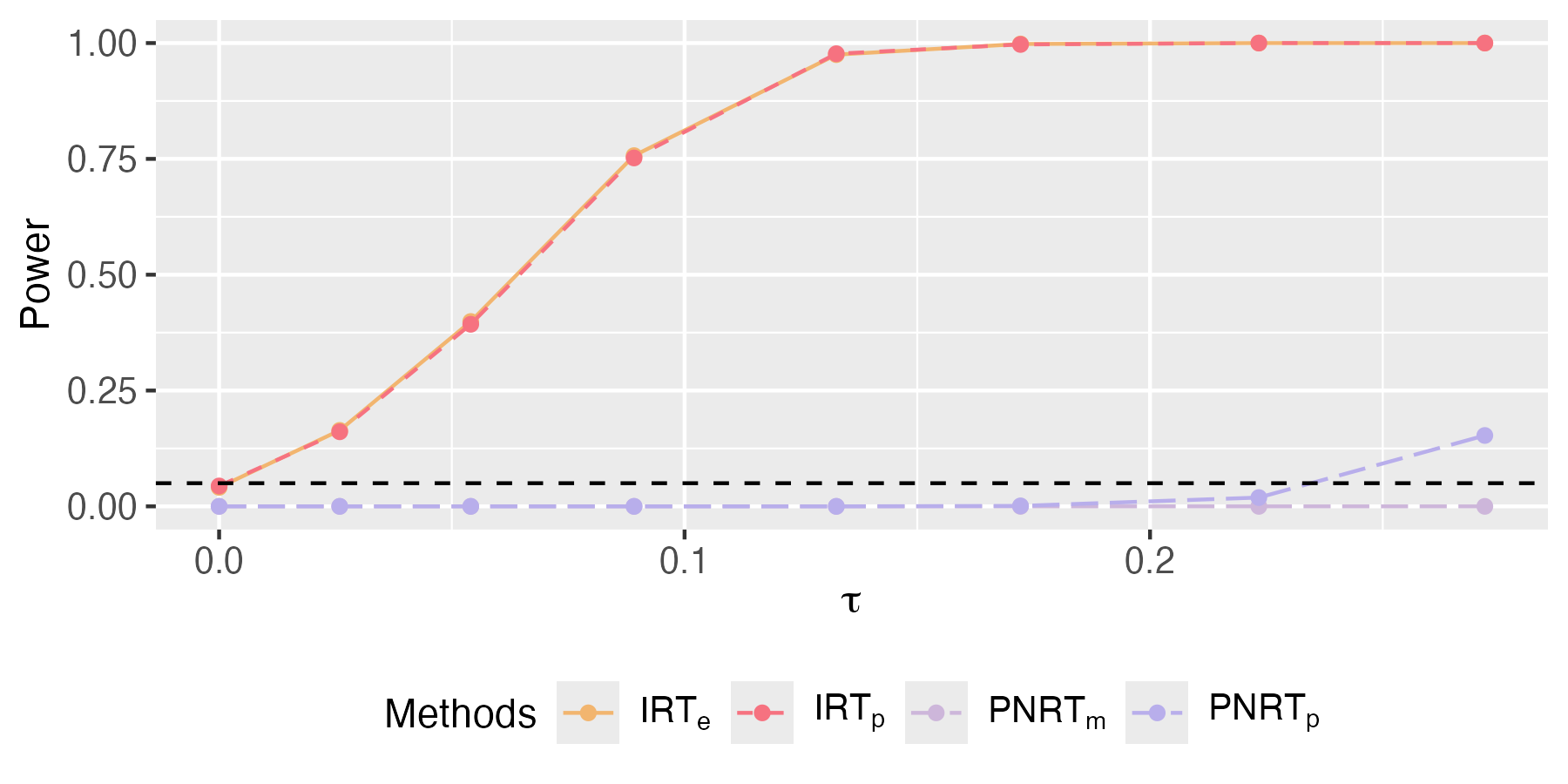}
\caption{Power of different randomization tests for simulated datasets generated from modified versions of a base model obtained by fitting the real data. The horizontal dashed line indicates the significance level of $\alpha  = 0.05$.}
\label{fig:plot_real}
\end{figure}

We apply PNRT$_\p$, PNRT$_\m$, IRT$_\e$, and IRT$_\p$ to this simulated dataset, and visualize their type I error rate (for cases with $\tau=0$) and power (for cases with $\tau>0$) in Fig. \ref{fig:plot_real}.
The results indicate that IRTs can effectively control the type I error rate and has high power to detect spillover effects,
while PNRTs have almost no power at all.

\section{Discussion}
\label{sec:summary}
In this paper, we proposed the IRT method, which imputes unknown nuisance parameters and computes the average of pristine \textit{p}-values. 
Theoretically, this method can bound the type I error rate at twice the significance level asymptotically when the missing rate of nuisance parameters converges to 0 in probability.
Simulation results show that even when the missing rate of nuisance parameters is relatively high, the IRT method demonstrates good empirical control of type I error rate.
Compared to CRTs and PNRTs, IRT methods significantly improve power and maintain robust performance regardless of the true distribution, particularly when the proportion of imputable outcomes under the null hypothesis is small.
Overall, we recommend IRT$_\e$, which imputes unknowns using the empirical distribution of the observed nuisance parameters, as it is easy to implement and free from the model specification.

Our IRT framework could be extended in the following directions.
First, covariates could be incorporated into IRT to further improve power, including using covariate information and machine learning methods to impute unobserved potential outcomes, and employing covariate-adjusted test statistics \citep{zhao2024adjust}.
Second, studentized test statistics could be used to construct asymptotic valid tests for weak null hypotheses \citep{wu2021randomization,cohen2022gaussian}.
Third, we could construct randomization-based confidence intervals by extending the analytical inversion approach of FRT to IRT \citep{zhu2023pair,fiksel2024exact}.
Finally, the IRT framework could be applied to other complex network experiments and complex interference structures \citep{basse2018model,leung2022rate,airoldi2024induction,liu2024cluster}.

\section{Funding}
This work was supported by the National Natural Science Foundation of China (12371269 \& 12071242).

\section*{Supplementary Material}\label{SM}
Supplementary Material available online includes proofs and additional simulation results.
An R package implementing the proposed methods is available at \href{https://github.com/htx113/imprt}{https://github.com/htx113/imprt}.

\FloatBarrier

\bibliographystyle{apalike}
\bibliography{paper-ref}

\newpage
\appendix

\begin{center}
    \textbf{\centering \Large Supplementary Material}
\end{center}
\setcounter{equation}{0}
\renewcommand{\theequation}{A\arabic{equation}}
\setcounter{figure}{0}
\renewcommand{\thefigure}{A\arabic{figure}}
\setcounter{lemma}{0}
\renewcommand{\thelemma}{A\arabic{lemma}}

Section~\ref{sec:proof} provides proofs of theoretical results. Section \ref{sec:sim} provides additional simulation results.

\section{Proofs}
\label{sec:proof}
\subsection{Proof of Theorem \ref{thm:oracle}}
By the law of iterated expectation,
we have 
\[\begin{split}
\bbE_{(\bZ^\obs,\btheta)}\left[h(p_{\MI}(\bZ^{\star};\btheta^{\obs}))\right]&= \bbE_{\bZ^\obs}\left[\bbE_{\btheta|\bZ^\obs}\left(h(p_{\MI}(\bZ^{\star};\btheta^{\obs}))\mid\bZ^{\star}\right)\right].
\end{split}\]
By definition, when oracle imputation distribution is known, 
\[\begin{split}
p_{\MI}(\bZ^{\star};\btheta^{\obs}) &= \int p_{\SI}(\bZ^{\star};\btheta^{\imp})\pi_o(\btheta^{\imp}\mid \btheta^{\obs}) d \btheta^{\imp}.
\end{split}\]
For any convex function $h$ on $[0,1]$, 
\[\begin{split}
h(p_{\MI}(\bZ^{\star};\btheta^{\obs})) &\leq \int h(p_{\SI}(\bZ^{\star};\btheta^{\imp}))\pi_o(\btheta^{\imp}\mid \btheta^{\obs}) d \btheta^{\imp}.
\end{split}\]
Since $h$ is bounded on $[0,1]$,  Fubini's theorem implies that
\[\begin{split}
\bbE_{\btheta|\bZ^\obs}\left(h(p_{\MI}(\bZ^{\star};\btheta^{\obs}))\mid \bZ^{\star}\right) &\leq \int \int h(p_{\SI}(\bZ^{\star};\btheta^{\imp}))\pi_o(\btheta^{\imp}\mid \btheta^{\obs}) d \btheta^{\imp}g(\btheta) d \btheta\\
&= \int \int h(p_{\SI}(\bZ^{\star};\btheta^{\imp}))\pi_o(\btheta^{\imp}\mid \btheta^{\obs}) g(\btheta)  d \btheta d \btheta^{\imp}.
\end{split}\]
By definition,
\[\begin{split}
\pi_o(\btheta^{\imp}\mid \btheta^\obs) &= \prod_{i\notin\bbU^{\star}_{a,b}} g(\theta_i^{\imp})\cdot \prod_{i\in\bbU^{\star}_{a,b}}\delta(\theta^{\imp}_{i} - \theta_i^\obs)\\
&= \prod_{i\notin\bbU^{\star}_{a,b}} g(\theta_i^{\imp})\cdot \prod_{i\in\bbU^{\star}_{a,b}}\delta(\theta^{\imp}_{i} - \theta_i).
\end{split}\]
Hence, 
conditional on $\bZ^\obs$,
\[\begin{split}
\int \pi_o(\btheta^{\imp}\mid \btheta^\obs)g(\btheta) d\btheta  &=\prod_{i\notin\bbU^{\star}_{a,b}} g(\theta_i^{\imp})\cdot \prod_{i\in\bbU^{\star}_{a,b}} \int 
\delta(\theta^{\imp}_{i} - \theta_i) g(\theta_i) d\theta_i\\&= \prod_{i=1}^N g(\theta_i^{\imp})
= g(\btheta^{\imp}), 
\end{split}\]
and we have 
\[\begin{split}
\bbE_{\btheta|\bZ^\obs}\left(h(p_{\MI}(\bZ^{\star};\btheta^{\obs}))\mid \bZ^{\star}\right)  &\leq \int h(p_{\SI}(\bZ^{\star};\btheta^{\imp}))g(\btheta^{\imp})  d \btheta^{\imp}.
\end{split}\]
Hence,  
\[\begin{split}
\bbE_{(\bZ^\obs,\btheta)}\left[h(p_{\MI}(\bZ^{\star};\btheta^{\obs}))\right]&= \bbE_{\bZ^\obs}\left[\bbE_{\btheta|\bZ^\obs}\left(h(p_{\MI}(\bZ^{\star};\btheta^{\obs}))\mid\bZ^{\star}\right)\right]\\
&\leq \int \int h(p_{\SI}(\bZ^{\star};\btheta^{\imp}))g(\btheta^{\imp})  d \btheta^{\imp} d\bF_\cZ(\bZ^{\star})\\
&= \int \int h(p_{\SI}(\bZ^{\star};\btheta^{\imp})) d\bF_\cZ(\bZ^{\star})g(\btheta^{\imp})d \btheta^{\imp} \\
&= \int \bbE_U[h(U)] g(\btheta^{\imp})d \btheta^{\imp} = \bbE_U[h(U)],
\end{split}\]
where $U \sim U[0,1]$.
Moreover, when $h(U) \equiv U$, we have 
\[\begin{split}
\bbE_{(\bZ^\obs,\btheta)}\left[h(p_{\MI}(\bZ^{\star};\btheta^{\obs}))\right]&=\bbE_U[U] = \frac{1}{2}.
\end{split}\]
We complete the proof.
\qed

\subsection{Proof of Corollary~\ref{cor:typeI}}
{We state Lemma 1 in \cite{meng1994posterior} in the following Lemma \ref{lem:meng}.
\begin{lemma}\label{lem:meng}
Let $G(\alpha)$ be the c.d.f. of a random variable $W$ on $[0,1]$. If $W$ is stochastically less variable than $U[0,1]$ but with the same mean, then $\forall \alpha \in [0,1]$,
\begin{equation}\label{eq:LemmaMeng}
\alpha - \left[\alpha^2-2\int_0^\alpha G(t) dt \right]^{1/2} \leq G(\alpha) \leq \alpha + \left[\alpha^2-2\int_0^\alpha G(t) dt \right]^{1/2} \leq 1.
\end{equation}
The first or second inequality becomes equality for all $\alpha$ if and only if $G(\alpha) \equiv \alpha$.
\end{lemma}

By Theorem \ref{thm:oracle}, the IRT$_\o$ \textit{p}-value $p_{\MI}(\bZ^{\star};\btheta^{\obs})$ is stochastically less variable than $U[0,1]$ but with the same mean when $\bZ^\obs \sim \bF_\cZ$ and $\theta_1,\ldots,\theta_N$ are i.i.d. with density $g(\cdot)$ known.
Denote the c.d.f. of $p_{\MI}(\bZ^{\star};\btheta^{\obs})$ as $G(\cdot)$.
Then Lemma \ref{lem:meng} gives the lower and upper bounds of $G(\alpha)$, and we have 
\[\begin{split}
\bbP(p_{\MI}(\bZ^{\star};\btheta^{\obs})\leq \alpha) = G(\alpha) \leq \alpha + \left[\alpha^2-2\int_0^\alpha G(t) dt \right]^{1/2} \leq 2\alpha,
\end{split}\]
where $\bbP$ is the probability taken with respect to (w.r.t.) $\bZ^\obs \sim \bF_\cZ$ and $\theta_1,\ldots,\theta_N$ i.i.d. with density $g(\cdot)$.
}
\qed

\subsection{Proof of Lemma \ref{lma:similar}}
Denote the \textit{p}-value of IRT$_\o$ as $p_{\MI}^o(\bZ^{\star};\btheta^{\obs})$, then for the \textit{p}-value $p_{\MI}(\bZ^{\star};\btheta^{\obs})$ of IRT with imputation function \eqref{eq:estimate}, we have 
\[\begin{split}
\Delta(\bZ^{\star};\btheta^{\obs}) = p_{\MI}(\bZ^{\star};\btheta^{\obs}) - p_{\MI}^o(\bZ^{\star};\btheta^{\obs}),
\end{split}\]
and 
\[\begin{split}
\bbP\left(p_{\MI}(\bZ^{\star};\btheta^{\obs}) \leq \alpha\right) &= \bbP\left(p_{\MI}^o(\bZ^{\star};\btheta^{\obs}) + \Delta(\bZ^{\star};\btheta^{\obs}) \leq \alpha\right)\\
&= \bbP\left(p_{\MI}^o(\bZ^{\star};\btheta^{\obs}) + \Delta(\bZ^{\star};\btheta^{\obs}) \leq \alpha, |\Delta(\bZ^{\star};\btheta^{\obs})| > \delta\right) \\
& \indent +\bbP\left(p_{\MI}^o(\bZ^{\star};\btheta^{\obs}) + \Delta(\bZ^{\star};\btheta^{\obs}) \leq \alpha, |\Delta(\bZ^{\star};\btheta^{\obs})| \leq \delta\right) \\
&\leq \bbP\left(|\Delta(\bZ^{\star};\btheta^{\obs})| > \delta\right) + \bbP\left(p_{\MI}^o(\bZ^{\star};\btheta^{\obs})  \leq \alpha + \delta\right)\\
&\leq \bbP\left(|\Delta(\bZ^{\star};\btheta^{\obs})| > \delta\right) + 2\alpha + 2\delta.
\end{split}\]
Since $|\Delta(\bZ^{\star};\btheta^{\obs})| = o_p(1)$, then for any $\delta > 0$, we have 
\[\begin{split}
\lim_{N\rightarrow\infty} \bbP\left(|\Delta(\bZ^{\star};\btheta^{\obs})| > \delta\right) = 0.
\end{split}\]
Therefore, 
\[\begin{split}
\limsup_{N\rightarrow\infty}
\bbP\left(p_{\MI}(\bZ^{\star};\btheta^{\obs}) \leq \alpha\right) \leq 2\alpha + 2\delta,\ \forall \delta > 0.
\end{split}\]
Letting $\delta \rightarrow 0$, we obtain 
\[\begin{split}
\limsup_{N\rightarrow\infty}
\bbP\left(p_{\MI}(\bZ^{\star};\btheta^{\obs}) \leq \alpha\right) \leq 2\alpha.
\end{split}\]
We complete the proof. 
\qed

\subsection{Proof of Theorem \ref{thm:SufficientCondition}}
By definition, $\btheta^{\obs}$ is determined by $(\btheta,\bZ^\obs)$.
In the following, we denote the expectation with respect to 
the joint distribution of $(\bZ^\obs,\btheta,\btheta^\imp)$ for $\bZ^\obs \sim \bF_\cZ$, $\theta_i \sim g(\cdot)$ i.i.d., and $(\btheta^\imp\mid\btheta,\bZ^\obs) \sim \pi_o(\btheta^{\imp}\mid \btheta^{\obs})$ as $\tilde{\bbE}$.

\hspace{\fill}\\
\noindent \textbf{(1) Continuous distribution:}
When $g(\cdot)$ and $\hat{g}(\cdot \mid \Tilde\btheta^\obs)$ correspond to continuous distributions, 
let 
\[\begin{split}
h(\btheta^{\imp}\mid\btheta^{\obs})
=\prod_{i\notin\bbU^{\star}_{a,b}}\frac{\hat{g}(\theta_i^{\imp}\mid\Tilde\btheta^{\obs})}{g(\theta_i^{\imp})}-1.
\end{split}\]
Then 
\[\begin{split}
\Delta(\bZ^{\star};\btheta^{\obs}) &= \int p_{\SI}(\bZ^{\star};\btheta^{\imp})h(\btheta^{\imp}\mid \btheta^{\obs}) \pi_o(\btheta^{\imp}\mid \btheta^{\obs})d \btheta^{\imp}.
\end{split}\]
The distribution of $\Delta(\bZ^{\star};\btheta^{\obs})$ depends on $(\bZ^{\obs},\btheta)$, and
it suffices to show that 
\[\begin{split}
\bbE_{(\bZ^\obs,\btheta)}|\Delta(\bZ^{\star};\btheta^{\obs})| \rightarrow 0,
\end{split}\]
where the expectation is taken with respect to the joint distribution of $(\bZ^{\obs},\btheta)$, and 
\[\begin{split}
|\Delta(\bZ^{\star};\btheta^{\obs})| \leq \int |h(\btheta^{\imp}\mid \btheta^{\obs})| \pi_o(\btheta^{\imp}\mid \btheta^{\obs})d \btheta^{\imp}.
\end{split}\]
Therefore,
\[\begin{split}
\bbE_{(\bZ^\obs,\btheta)}|\Delta(\bZ^{\star};\btheta^{\obs})| &\leq \int\int\int |h(\btheta^{\imp}\mid \btheta^{\obs})| \pi_o(\btheta^{\imp}\mid \btheta^{\obs})d \btheta^{\imp}g(\btheta) d\btheta d\bF_\cZ(\bZ^\obs)\\
&= \tilde{\bbE}|h(\btheta^{\imp}\mid \btheta^{\obs})|,
\end{split}\]
where $\tilde{E}$ denotes the expectation with respect to the joint distribution of $(\bZ^\obs,\btheta,\btheta^\imp)$ for $\bZ^\obs \sim \bF_\cZ$, $\theta_i \sim g(\cdot)$ i.i.d., and $(\btheta^\imp\mid\btheta,\bZ^\obs) \sim \pi_o(\btheta^{\imp}\mid \btheta^{\obs})$.
Therefore, a sufficient condition for $\Delta(\bZ^{\star};\btheta^{\obs}) = o_p(1)$ is
\begin{equation}\label{eq:target1}
\tilde{\bbE}|h(\btheta^{\imp}\mid \btheta^{\obs})| \rightarrow 0.
\end{equation}
By definition,
\[\begin{split}
\pi_o(\btheta^{\imp}\mid \btheta^\obs) &= \prod_{i\notin\bbU^{\star}_{a,b}} g(\theta_i^{\imp})\cdot \prod_{i\in\bbU^{\star}_{a,b}}\delta(\theta^{\imp}_{i} - \theta_i^\obs)\\
&= \prod_{i\notin\bbU^{\star}_{a,b}} g(\theta_i^{\imp})\cdot \prod_{i\in\bbU^{\star}_{a,b}}\delta(\theta^{\imp}_{i} - \theta_i),
\end{split}\]
and $\tilde{\btheta}^\obs = \{\theta_i^\obs: i \in \bbU^{\star}_{a,b}\} \equiv \{\theta_i: i \in \bbU^{\star}_{a,b}\}$.
For a given $\bZ^\obs$, the set $\bbU_{a,b}^\obs$ is determined. 
Denote $\bbU_{a,b}^\obs = \{i_1,\ldots,i_{N_{a,b}^\obs}\}$ and $(\bbU_{a,b}^\obs)^c = \{i_{N_{a,b}^\obs+1},\ldots,i_{N}\}$, and their union is $\bbU = \{1,\ldots,N\}$.
Then the conditional expectation of $|h(\btheta^{\imp}\mid \btheta^{\obs})|$ given $\bZ^\obs$ is 
\[\begin{split}
&\indent {\bbE}\left(|h(\btheta^{\imp}\mid \btheta^{\obs})|\mid \bZ^\obs\right) \\ &= \int\int |h(\btheta^{\imp}\mid \btheta^{\obs})| \pi_o(\btheta^{\imp}\mid \btheta^{\obs})g(\btheta)d \btheta^{\imp} d\btheta \\
&= \int\int\left|\prod_{i\notin\bbU^{\star}_{a,b}}\frac{\hat{g}(\theta_i^{\imp}\mid\Tilde\btheta^{\obs})}{g(\theta_i^{\imp})}-1\right|\prod_{i\notin\bbU^{\star}_{a,b}}g(\theta_i^{\imp}) \prod_{i\in\bbU^{\star}_{a,b}}
\delta(\theta_i^\imp - \theta_i)\cdot 
\prod_{i=1}^N
g(\theta_i)
d\btheta^\imp d\btheta \\
&= \int\int\left|\prod_{i\notin\bbU^{\star}_{a,b}}\frac{\hat{g}(\theta_i^{\imp}\mid\Tilde\btheta^{\obs})}{g(\theta_i^{\imp})}-1\right|\prod_{i\notin\bbU^{\star}_{a,b}}g(\theta_i^{\imp}) \prod_{i\in\bbU^{\star}_{a,b}}
g(\theta_i)
d\btheta^\imp_{i\notin\bbU^{\star}_{a,b}} d\btheta_{i\in\bbU^{\star}_{a,b}} \\
&= \int\int\left|\prod_{k = N_{a,b}^\obs + 1}^N\frac{\hat{g}\left(\theta_{i_k}^{\imp}\mid \btheta_{i_1:i_{N_{a,b}^\obs}}\right)}{g(\theta_{i_k}^{\imp})}-1\right|\prod_{k=N_{a,b}^\obs + 1}^{N} g(\theta_{i_k}^{\imp}) \prod_{j=1}^{N_{a,b}^\obs}
g(\theta_{i_j})
d\btheta^\imp_{i_{N_{a,b}^\obs+1}:i_N} d\btheta_{i_1:i_{N_{a,b}^\obs}}.
\end{split}\]
We make the following variable substitution: $\vt_k = \theta_{i_k}$ for $k = 1,\ldots,N_{a,b}^\obs$, and $\vt_k = \theta_{i_k}^\imp$ for $k = N_{a,b}^\obs+1,\ldots,N$.
Then we have 
\[\begin{split} {\bbE}\left(|h(\btheta^{\imp}\mid \btheta^{\obs})|\mid \bZ^\obs\right) = \int\left|\prod_{k = N_{a,b}^\obs + 1}^N\frac{\hat{g}\left(\vt_k\mid \bmvt_{1:N_{a,b}^\obs}\right)}{g(\vt_k)}-1\right|\prod_{k=1}^{N} g(\vt_k)
d \bmvt_{1:N},
\end{split}\]
which is equal to the conditional expectation of 
\[\begin{split}
\left|\prod_{k=N_{a,b}^\obs+1}^{N} \frac{\hat{g}\left(\vt_k \mid \bmvt_{1:N_{a,b}^\obs}\right)}{g(\vt_k)}-1\right|
\end{split}\]
with respect to $\vt_1,\ldots,\vt_N$ i.i.d. with probability density $g(\cdot)$ given $N_{a,b}^\obs$.
Hence,
\[\begin{split}
&\indent\tilde{\bbE}|h(\btheta^{\imp}\mid \btheta^{\obs})|
= \bbE\left[{\bbE}\left(|h(\btheta^{\imp}\mid \btheta^{\obs})|\mid \bZ^\obs\right)\right]\\
&= \sum_{\bz \in \cZ}{\bbE}\left(|h(\btheta^{\imp}\mid \btheta^{\obs})|\mid \bZ^\obs = \bz\right)\bbP_\bZ(\bZ=\bz)\\
&= \sum_{\bz \in \cZ}{\bbE}\left(\left|\prod_{k = N_{a,b}^\obs + 1}^N\frac{\hat{g}\left(\vt_k\mid \bmvt_{1:N_{a,b}^\obs}\right)}{g(\vt_k)}-1\right|\mid N_{a,b}^\obs = |\bbU_a(\bz)\cup \bbU_b(\bz)|\right)\bbP_\bZ(\bZ=\bz)\\
&= \sum_{n}\bbE\left(\left|\prod_{k = N_{a,b}^\obs + 1}^N\frac{\hat{g}\left(\vt_k\mid \bmvt_{1:N_{a,b}^\obs}\right)}{g(\vt_k)}-1\right|\mid N_{a,b}^\obs=n \right)\sum_{\bz: |\bbU_a(\bz)\cup \bbU_b(\bz)|=n}\bbP_\bZ(\bZ=\bz)\\
&=
\sum_{n}\bbE\left(\left|\prod_{k = N_{a,b}^\obs + 1}^N\frac{\hat{g}\left(\vt_k\mid \bmvt_{1:N_{a,b}^\obs}\right)}{g(\vt_k)}-1\right|\mid N_{a,b}^\obs=n \right)\bbP_N(N_{a,b}^\obs=n)\\
&= \bbE\left|\prod_{k = N_{a,b}^\obs + 1}^N\frac{\hat{g}\left(\vt_k\mid \bmvt_{1:N_{a,b}^\obs}\right)}{g(\vt_k)}-1\right|,
\end{split}\]
where $\bbP_N(N_{a,b}^\obs=n) = \sum_{\bz: |\bbU_a(\bz)\cup \bbU_b(\bz)|=n}\bbP_\bZ(\bZ=\bz)$ is the p.m.f. of $\bF_N$, and 
the last expectation is with respect to $\vt_1,\ldots,\vt_N$ i.i.d. with probability density $g(\cdot)$, and $N_{a,b}^\obs \sim \bF_N$.
Thus, sufficient condition \eqref{eq:target1} is equivalent to:
\begin{equation}
\label{eq:SufficientCondition_last}\bbE\left|\prod_{k = N_{a,b}^\obs + 1}^N\frac{\hat{g}\left(\vt_k\mid \bmvt_{1:N_{a,b}^\obs}\right)}{g(\vt_k)}-1\right| \rightarrow 0,
\end{equation}
where the expectation is taken with respect to  $\vt_1,\ldots,\vt_N$ i.i.d. with probability density $g(\cdot)$ and $N_{a,b}^\obs\sim \bF_N$.

\hspace{\fill}\\
\noindent \textbf{(2) Discrete distribution:}
When $g(\cdot)$ and $\hat{g}(\cdot \mid \Tilde\btheta^\obs)$ correspond to discrete distributions, we denote their corresponding probability mass function as $p(\cdot)$ and $\hat{p}(\cdot \mid \Tilde\btheta^\obs)$, i.e.,
\[\begin{split}
g(\theta_i^\imp) = \sum_{\psi \in \Psi} p(\psi)\delta(\theta_i^\imp - \psi),\quad 
\hat{g}(\theta_i^\imp \mid \Tilde\btheta^\obs) = \sum_{\psi \in \Psi} \hat{p}(\psi \mid \Tilde\btheta^\obs)\delta(\theta_i^\imp - \psi),
\end{split}\]
where $\Psi$ is the set of all the $\psi$ s.t. $p(\psi) > 0$.
Then for any function $\nu(\btheta^{\imp} \mid \btheta^\obs)$, 
\[\begin{split}
&\indent \int \nu(\btheta^{\imp} \mid \btheta^\obs) \pi_o(\btheta^{\imp}\mid \btheta^{\obs}) d\btheta^\imp\\ &=
\int \nu(\btheta^{\imp} \mid \btheta^\obs) \cdot\prod_{i \notin \bbU_{a,b}^\obs}  \sum_{\psi \in \Psi} p(\psi)\delta(\theta_i^\imp - \psi)\cdot 
\prod_{i \in \bbU_{a,b}^\obs} \delta(\theta_i^\imp - \theta_i^\obs)
d\btheta^\imp\\
&= \int \nu(\btheta^{\imp} \mid \btheta^\obs) \cdot \sum_{\psi_i \in \Psi \text{ for } i \notin \bbU_{a,b}^\obs} \prod_{i \notin \bbU_{a,b}^\obs} p(\psi_i)\delta(\theta_i^\imp - \psi_i)\cdot 
\prod_{i \in \bbU_{a,b}^\obs} \delta(\theta_i^\imp - \theta_i^\obs)
d\btheta^\imp\\
&= \sum_{\psi_i \in \Psi \text{ for } i \notin \bbU_{a,b}^\obs} \int \nu(\btheta^{\imp} \mid \btheta^\obs) \prod_{i \notin \bbU_{a,b}^\obs} p(\psi_i) \cdot \prod_{i \notin \bbU_{a,b}^\obs} \delta(\theta_i^\imp - \psi_i)\cdot 
\prod_{i \in \bbU_{a,b}^\obs} \delta(\theta_i^\imp - \theta_i^\obs)
d\btheta^\imp\\
&= \sum_{\btheta^\imp \in \Theta^\obs} \nu(\btheta^{\imp} \mid \btheta^\obs) \prod_{i \notin \bbU_{a,b}^\obs} p(\theta_i^\imp),
\end{split}\]
where $\Theta^\obs = \{\btheta^\imp: \theta_i^\imp = \theta_i^\obs\text{ for }i \in \bbU_{a,b}^\obs,
\text{ and }
\theta_i^\imp \in \Psi \text{ for }i \notin \bbU_{a,b}^\obs\}$. Similarly, we have 
\[\begin{split}
\int \nu(\btheta^{\imp}\mid \btheta^\obs) \hat{\pi}(\btheta^{\imp}\mid \btheta^{\obs}) d\btheta^\imp = \sum_{\btheta^\imp \in \Theta^\obs} \nu(\btheta^{\imp}\mid \btheta^\obs) \prod_{i \notin \bbU_{a,b}^\obs} \hat{p}(\theta_i^\imp \mid \Tilde\btheta^\obs).
\end{split}\]
Hence,
\[\begin{split}
\Delta(\bZ^{\star};\btheta^{\obs})
&=\int p_{\SI}(\bZ^{\star};\btheta^{\imp})\left[\hat{\pi}(\btheta^{\imp}\mid \btheta^{\obs})-\pi_o(\btheta^{\imp}\mid \btheta^{\obs})\right] d \btheta^{\imp}\\
&= \sum_{\btheta^\imp \in \Theta^\obs}p_{\SI}(\bZ^{\star};\btheta^{\imp}) \prod_{i \notin \bbU_{a,b}^\obs} \hat{p}(\theta_i^\imp \mid \Tilde\btheta^\obs)
- \sum_{\btheta^\imp \in \Theta^\obs}p_{\SI}(\bZ^{\star};\btheta^{\imp}) \prod_{i \notin \bbU_{a,b}^\obs} p(\theta_i^\imp)\\
&=\sum_{\btheta^\imp \in \Theta^\obs}p_{\SI}(\bZ^{\star};\btheta^{\imp}) \prod_{i \notin \bbU_{a,b}^\obs} p(\theta_i^\imp)
\left(\prod_{i \notin \bbU_{a,b}^\obs} \frac{\hat{p}(\theta_i^\imp \mid \Tilde\btheta^\obs)}{p(\theta_i^\imp)}-1\right),
\end{split}\]
where $\Theta^\obs = \{\btheta^\imp: \theta_i^\imp = \theta_i^\obs\text{ for }i \in \bbU_{a,b}^\obs,
\text{ and }
\theta_i^\imp \in \Psi \text{ for }i \notin \bbU_{a,b}^\obs\}$.
Let 
\[\begin{split}
h_d(\btheta^{\imp}\mid\btheta^{\obs})=
\prod_{i\notin\bbU^{\star}_{a,b}}\frac{\hat{p}(\theta_i^{\imp}\mid\Tilde\btheta^{\obs})}{p(\theta_i^{\imp})}-1,
\end{split}\]
then 
\begin{equation}\label{eq:target_discrete}
\Delta(\bZ^{\star};\btheta^{\obs}) = \int p_{\SI}(\bZ^{\star};\btheta^{\imp})h_d(\btheta^{\imp}\mid \btheta^{\obs}) 
\pi_o(\btheta^{\imp}\mid \btheta^{\obs})d \btheta^{\imp}.
\end{equation}
Similar to the case of continuous distribution, the target is
\[\begin{split}
\bbE_{(\bZ^\obs,\btheta)}|\Delta(\bZ^{\star};\btheta^{\obs})| \rightarrow 0,
\end{split}\]
where the expectation is taken with respect to the joint distribution of $(\bZ^{\obs},\btheta)$, and 
\[\begin{split}
|\Delta(\bZ^{\star};\btheta^{\obs})| \leq \int |h_d(\btheta^{\imp}\mid \btheta^{\obs})| \pi_o(\btheta^{\imp}\mid \btheta^{\obs})d \btheta^{\imp}.
\end{split}\]
Therefore,
\[\begin{split}
\bbE_{(\bZ^\obs,\btheta)}|\Delta(\bZ^{\star};\btheta^{\obs})| &\leq \int\int\int |h_d(\btheta^{\imp}\mid \btheta^{\obs})| \pi_o(\btheta^{\imp}\mid \btheta^{\obs})d \btheta^{\imp}g(\btheta) d\btheta d\bF_\cZ(\bZ^\obs)\\
&= \tilde{\bbE}|h_d(\btheta^{\imp}\mid \btheta^{\obs})|,
\end{split}\]
and the following is a sufficient condition for $\Delta(\bZ^{\star};\btheta^{\obs}) = o_p(1)$: 
\begin{equation}\label{eq:target2}
\tilde{\bbE}|h_d(\btheta^{\imp}\mid \btheta^{\obs})| \rightarrow 0.
\end{equation}
Similarly, the conditional expectation of $|h_d(\btheta^{\imp}\mid \btheta^{\obs})|$ given $\bZ^\obs$ is 
\[\begin{split}
&\indent {\bbE}\left(|h_d(\btheta^{\imp}\mid \btheta^{\obs})|\mid \bZ^\obs\right) \\ &= \int\int |h_d(\btheta^{\imp}\mid \btheta^{\obs})| \pi_o(\btheta^{\imp}\mid \btheta^{\obs})g(\btheta)d \btheta^{\imp} d\btheta \\
&= \int\int\left|\prod_{i\notin\bbU^{\star}_{a,b}}\frac{\hat{p}(\theta_i^{\imp}\mid\Tilde\btheta^{\obs})}{p(\theta_i^{\imp})}-1\right|\prod_{i\notin\bbU^{\star}_{a,b}}g(\theta_i^{\imp}) \prod_{i\in\bbU^{\star}_{a,b}}
\delta(\theta_i^\imp - \theta_i)\cdot 
\prod_{i=1}^N
g(\theta_i)
d\btheta^\imp d\btheta \\
&= \int\int\left|\prod_{i\notin\bbU^{\star}_{a,b}}\frac{\hat{p}(\theta_i^{\imp}\mid\Tilde\btheta^{\obs})}{p(\theta_i^{\imp})}-1\right|\prod_{i\notin\bbU^{\star}_{a,b}}g(\theta_i^{\imp}) \prod_{i\in\bbU^{\star}_{a,b}}
g(\theta_i)
d\btheta^\imp_{i\notin\bbU^{\star}_{a,b}} d\btheta_{i\in\bbU^{\star}_{a,b}} \\
&= \int\left|\prod_{k=N_{a,b}^\obs+1}^{N}\frac{\hat{p}\left(\vt_k\mid \bmvt_{1:{N_{a,b}^\obs}}\right)}{p(\vt_k)}-1\right|\prod_{k=1}^{N}g(\vt_k) d \bmvt_{1:N},
\end{split}\]
which is equal to the conditional expectation of 
\[\begin{split}
\left|\prod_{k=N_{a,b}^\obs+1}^{N}\frac{\hat{p}\left(\vt_k\mid \bmvt_{1:{N_{a,b}^\obs}}\right)}{p(\vt_k)}-1\right|
\end{split}\]
with respect to $\vt_1,\ldots,\vt_N$ i.i.d. with probability density $g(\cdot)$ given $N_{a,b}^\obs$.
Hence, we can similarly prove that 
\[\begin{split}
\tilde{\bbE}|h_d(\btheta^{\imp}\mid \btheta^{\obs})|
&= \bbE\left[{\bbE}\left(|h_d(\btheta^{\imp}\mid \btheta^{\obs})|\mid \bZ^\obs\right)\right]\\
&= \bbE\left|\prod_{k=N_{a,b}^\obs+1}^{N}\frac{\hat{p}\left(\vt_k\mid \bmvt_{1:{N_{a,b}^\obs}}\right)}{p(\vt_k)}-1\right|,
\end{split}\]
where the last expectation is with respect to $\vt_1,\ldots,\vt_N$ i.i.d. with probability density $g(\cdot)$, and $N_{a,b}^\obs \sim \bF_N$.
Thus, sufficient condition \eqref{eq:target2} is equivalent to:
\begin{equation}
\label{eq:SufficientConditionDiscrete2_last}\bbE\left|\prod_{k=N_{a,b}^\obs+1}^{N}\frac{\hat{p}\left(\vt_k\mid \bmvt_{1:{N_{a,b}^\obs}}\right)}{p(\vt_k)}-1\right| \rightarrow 0,
\end{equation}
where the expectation is taken with respect to $\vt_1,\ldots,\vt_N$ i.i.d. with probability density $g(\cdot)$ and $N_{a,b}^\obs\sim \bF_N$. 
We complete the proof.
\qed

\subsection{Proof of Corollary \ref{cor:normal_example}}\label{sec:normal_known}

Denote $N_1 = N_{a,b}^\obs = N (1-r_N^\mis)$ and $N_0 = N-N_{a,b}^\obs = N\cdot r_N^\mis$.
Suppose $\vt_1,\ldots,\vt_{N_1}\sim \cN(\mu,\sigma^2)$ i.i.d.  with $\sigma^2$ known, and conjugate prior $\mu \sim \cN(\mu_0,\sigma_0^2)$.
Let $g$ be the p.d.f. of $\cN(\mu,\sigma^2)$.
{ For $\bmvt_{1:N_1} = (\vt_1,\ldots,\vt_{N_1})$, define the average as $\bar{\bmvt}_{1:N_1} = N_1^{-1}\sum_{i=1}^{N_1} \vt_i$.}
Then the posterior distribution of $\mu$ is $\mu \mid \bmvt_{1:N_1} \sim \cN\left(\mu_{N_1},\sigma_{N_1}^2\right)$,
where $$\sigma_{N_1}^2 = \frac{1}{\frac{1}{\sigma_0^2}+\frac{N_1}{\sigma^2}},\quad \mu_{N_1} = \frac{\frac{\mu_0}{\sigma_0^2}+\frac{N_1\bar{\bmvt}_{1:N_1}}{\sigma^2}}{\frac{1}{\sigma_0^2}+\frac{N_1}{\sigma^2}}
= \bar{\bmvt}_{1:N_1} - \frac{\frac{\bar{\bmvt}_{1:N_1}-\mu_0}{\sigma_0^2}}{\frac{1}{\sigma_0^2}+\frac{N_1}{\sigma^2}},$$
and the posterior predictive distribution of $\theta_i$ is 
\[\begin{split}
\vt\mid \bmvt_{1:N_1} \sim \cN\left(\mu_{N_1},\sigma_{N_1}^2+\sigma^2\right).
\end{split}\]
Hence, for $i = N_1 + 1, \ldots, N$, 
\[\begin{split}
\hat{g}(\vt_i\mid\bmvt_{1:N_1}) = \frac{1}{\sqrt{2\pi(\sigma_{N_1}^2+\sigma^2)}}
\exp\left\{-\frac{(\vt_i-\mu_{N_1})^2}{2(\sigma_{N_1}^2+\sigma^2)}\right\},
\end{split}\]
and 
\[\begin{split}
\frac{\hat{g}(\vt_i\mid\bmvt_{1:N_1})}{g(\vt_i)} &=
\left(\frac{\sigma^2}{\sigma_{N_1}^2+\sigma^2}\right)^{1/2}
\exp\left\{\frac{(\vt_i-\mu)^2}{2\sigma^2}-\frac{(\vt_i-\mu_{N_1})^2}{2(\sigma_{N_1}^2+\sigma^2)}\right\}.
\end{split}\]
It suffices to show that 
\[\begin{split}
\bbE \left(\prod_{i=1}^{N_0}\frac{\hat{g}\left(\vt_i\mid \bmvt_{1:N_1}\right)}{g(\vt_i)}\right)^2 \rightarrow 0\text{ as }N\rightarrow\infty.
\end{split}\]

Given $\bar\bmvt_{1:N_1}$, $\hat{g}\left(\vt_i\mid \bmvt_{1:N_1}\right)/g(\vt_i)$ are i.i.d. for $i=N_1+1,\ldots,N$. 
Since $\bF_N$ is a point mass, then $N_1$ and $N_0$ are deterministic for each $N$, and 
\[\begin{split}
\bbE \left[\left(\prod_{i=N_1+1}^{N}\frac{\hat{g}\left(\vt_i\mid \bmvt_{1:N_1}\right)}{g(\vt_i)}\right)^2 \mid \bar\bmvt_{1:N_1}\right] = \left(\bbE\left[\left(\frac{\hat{g}\left(\vt_i\mid \bmvt_{1:N_1}\right)}{g(\vt_i)}\right)^2 \mid \bar\bmvt_{1:N_1}\right]\right)^{N_0},
\end{split}\]
where 
\[\begin{split}
\bbE\left[\left(\frac{\hat{g}\left(\vt_i\mid \bmvt_{1:N_1}\right)}{g(\vt_i)}\right)^2 \mid \bar\bmvt_{1:N_1}\right] &= \int \frac{\hat{g}^2\left(\vt\mid \bmvt_{1:N_1}\right)}{g(\vt)} d\vt\\
&= \frac{\sigma}{\sqrt{2\pi}(\sigma_{N_1}^2 + \sigma^2)} \int \exp\left\{\frac{(\vt-\mu)^2}{2\sigma^2} - \frac{(\vt-\mu_{N_1})^2}{\sigma_{N_1}^2 + \sigma^2}\right\} d\vt\\
&= \left(\frac{\sigma^4}{\sigma^4-\sigma_{N_1}^4}\right)^{1/2}\exp\left\{\frac{(\mu_{N_1}-\mu)^2}{\sigma^2-\sigma^2_{N_1}}\right\}.
\end{split}\]
Hence,
\[\begin{split}
\bbE \left(\prod_{i=1}^{N_0}\frac{\hat{g}\left(\vt_i\mid \bmvt_{1:N_1}\right)}{g(\vt_i)}\right)^2
&= \left(\frac{\sigma^4}{\sigma^4-\sigma_{N_1}^4}\right)^{N_0/2} \bbE_{\bar\bmvt_{1:N_1}}\left[
\exp\left\{\frac{N_0(\mu_{N_1}-\mu)^2}{\sigma^2-\sigma^2_{N_1}}\right\}\right]\\
&= \left(\frac{\sigma^4}{\sigma^4-\sigma_{N_1}^4}\right)^{N_0/2}\frac{\sqrt{N_1}}{\sqrt{2\pi}\sigma} \int \exp\left\{\frac{N_0(\mu_{N_1}-\mu)^2}{\sigma^2-\sigma^2_{N_1}} - \frac{N_1(\bar\bmvt_{1:N_1} - \mu)^2}{2\sigma^2}\right\} d\bar\bmvt_{1:N_1}.
\end{split}\]
As $r_N^\mis \rightarrow 0$ implies that  $N_0/N_1 \rightarrow 0$ and $N_1 \rightarrow \infty$, then there exists $\tilde{N} > 1$ s.t. for any $N > \tilde{N}$, we have $N_0/N_1 \leq 1/8$, $\sigma_{N_1}^2/\sigma^2 \leq 1/2$ and  $\sigma_{N_1}^2/\sigma^2_0 \leq 1/2$.
Let $a_N = \sigma^2_{N_1}/\sigma^2_0$,
then 
\[\begin{split}
\mu_{N_1}-\mu = (\bar{\bmvt}_{1:N_1}-\mu)(1-a_N) + (\mu_0-\mu)a_N,
\end{split}\]
and 
\[\begin{split}
&\indent\frac{N_0(\mu_{N_1}-\mu)^2}{\sigma^2-\sigma^2_{N_1}} - \frac{N_1(\bar\bmvt_{1:N_1} - \mu)^2}{2\sigma^2} \\
&= -\frac{N_1}{2\sigma^2}\left[1-\frac{2N_0\sigma^2(1-a_N)^2}{N_1(\sigma^2-\sigma^2_{N_1})}\right](\bar{\bmvt}_{1:N_1}-\mu)^2 + 
\frac{2N_0a_N(1-a_N)(\mu_0-\mu)}{\sigma^2-\sigma^2_{N_1}}(\bar{\bmvt}_{1:N_1}-\mu) + \frac{N_0(\mu_0-\mu)^2a_N^2}{\sigma^2-\sigma^2_{N_1}}.
\end{split}\]
Let 
\[\begin{split}
b_N &= 1-\frac{2N_0\sigma^2(1-a_N)^2}{N_1(\sigma^2-\sigma^2_{N_1})},\\
c_N &= \frac{2\sigma^2\cdot N_0a_N(1-a_N)(\mu_0-\mu)}{N_1(\sigma^2-\sigma^2_{N_1})},\\
d_N &= \frac{N_0(\mu_0-\mu)^2a_N^2}{\sigma^2-\sigma^2_{N_1}},
\end{split}\]
then for $N > \tilde{N}$, we have $0 \leq a_N \leq 1/2$,
$1/2 \leq b_N \leq 1$, and
\[\begin{split}
&\indent\frac{N_0(\mu_{N_1}-\mu)^2}{\sigma^2-\sigma^2_{N_1}} - \frac{N_1(\bar\bmvt_{1:N_1} - \mu)^2}{2\sigma^2} \\
&= -\frac{N_1}{2\sigma^2}\left[b_N(\bar{\bmvt}_{1:N_1}-\mu)^2 - 2c_N(\bar{\bmvt}_{1:N_1}-\mu)\right] + d_N \\
&= -\frac{N_1b_N}{2\sigma^2}\left[(\bar{\bmvt}_{1:N_1}-\mu)^2 - \frac{2c_N}{b_N}(\bar{\bmvt}_{1:N_1}-\mu)\right] + d_N\\
&= -\frac{N_1b_N}{2\sigma^2}\left[(\bar{\bmvt}_{1:N_1}-\mu) - \frac{c_N}{b_N}\right]^2 + d_N + \frac{N_1c_N^2}{2b_N\sigma^2}.
\end{split}\]
Hence, for $N > \tilde{N}$,
\[\begin{split}
\frac{\sqrt{N_1}}{\sqrt{2\pi}\sigma} \int \exp\left\{\frac{N_0(\mu_{N_1}-\mu)^2}{\sigma^2-\sigma^2_{N_1}} - \frac{N_1(\bar\bmvt_{1:N_1} - \mu)^2}{2\sigma^2}\right\} d\bar\bmvt_{1:N_1}= \frac{1}{\sqrt{b_N}}\exp\left\{d_N + \frac{N_1c_N^2}{2b_N\sigma^2}\right\},
\end{split}\]
and 
\[\begin{split}
\bbE \left(\prod_{i=N_1+1}^{N}\frac{\hat{g}\left(\vt_i\mid \bmvt_{1:N_1}\right)}{g(\vt_i)}\right)^2 = \left(\frac{\sigma^4}{\sigma^4-\sigma_{N_1}^4}\right)^{N_0/2}\frac{1}{\sqrt{b_N}}\exp\left\{d_N + \frac{N_1c_N^2}{2b_N\sigma^2}\right\}.
\end{split}\]
As $\sigma^2_{N_1}/\sigma^2 = 1/(N_1 + \sigma^2/\sigma_0^2) \leq 1/N_1$, then
\[\begin{split}
1\leq \left(\frac{\sigma^4}{\sigma^4-\sigma_{N_1}^4}\right)^{N_0/2} &= \left(1 + \frac{\sigma^4_{N_1}}{\sigma^4-\sigma_{N_1}^4}\right)^{N_0/2}= \left(1 + \frac{\sigma^4_{N_1}/\sigma^4}{1-\sigma_{N_1}^4/\sigma^4}\right)^{N_0/2}\\
&\leq \left(1 + \frac{1/N_1^2}{1-1/N_1^2}\right)^{N_0/2}
\leq \left(1 + \frac{2}{N_1^2}\right)^{N_0/2}\\
&= \left[\left(1 + \frac{2}{N_1^2}\right)^{N_1^2/2}\right]^{N_0/N_1^2}
\leq e^{N_0/N_1^2} \rightarrow 1,
\end{split}\]
i.e., 
\[\begin{split}
\lim_{N\rightarrow\infty} \left(\frac{\sigma^4}{\sigma^4-\sigma_{N_1}^4}\right)^{N_0/2} = 1.
\end{split}\]
Since $\lim_{N\rightarrow\infty} b_N = 1$, 
\[\begin{split}
\lim_{N\rightarrow\infty} N_0 a_N^2 &= \lim_{N\rightarrow\infty} \frac{N_0\sigma^4}{(N_1\sigma_0^2 + \sigma^2)^2} = 0,\\
\lim_{N\rightarrow\infty} N_1 c_N^2 &= \lim_{N\rightarrow\infty} \frac{4\sigma^4}{(\sigma^2-\sigma_{N_1}^2)^2}\cdot \frac{N_0}{N_1}\cdot N_0a_N^2(1-a_N)^2 = 0,
\end{split}\]
we have $\lim_{N\rightarrow\infty} d_N = 0$ and 
\[\begin{split}
\lim_{N\rightarrow\infty} \bbE \left(\prod_{i=N_1+1}^{N}\frac{\hat{g}\left(\vt_i\mid \bmvt_{1:N_1}\right)}{g(\vt_i)}\right)^2 = 1.
\end{split}\]
{Moreover, 
\[\begin{split}
\bbE \left(\prod_{i=N_1+1}^{N}\frac{\hat{g}\left(\vt_i\mid \bmvt_{1:N_1}\right)}{g(\vt_i)}\right) &= \int \prod_{i=N_1+1}^{N}\frac{\hat{g}\left(\vt_i\mid \bmvt_{1:N_1}\right)}{g(\vt_i)} \cdot \prod_{i=1}^{N} g(\vt_i) d\bmvt\\
&= \int \prod_{i=N_1+1}^{N}\hat{g}\left(\vt_i\mid \bmvt_{1:N_1}\right) \cdot \prod_{i=1}^{N_1} g(\vt_i) d\bmvt \equiv 1.
\end{split}\]}
Therefore, 
\[\begin{split}
\lim_{N\rightarrow\infty}\bbE \left(\prod_{i=N_1+1}^{N}\frac{\hat{g}\left(\vt_i\mid \bmvt_{1:N_1}\right)}{g(\vt_i)}-1\right)^2
&= \lim_{N\rightarrow\infty}\bbE \left(\prod_{i=N_1+1}^{N}\frac{\hat{g}\left(\vt_i\mid \bmvt_{1:N_1}\right)}{g(\vt_i)}\right)^2 -1 = 0,
\end{split}\]
and thus,
\[\begin{split}
\bbE \left|\prod_{i=N_1+1}^{N}\frac{\hat{g}\left(\vt_i\mid \bmvt_{1:N_1}\right)}{g(\vt_i)}-1\right| = o(1).
\end{split}\]
We complete the proof.
\qed

\subsection{Numerical verification of sufficient conditions in Theorem \ref{thm:SufficientCondition}}

For complex cases, theoretically verifying the sufficient conditions outlined in Theorem \ref{thm:SufficientCondition} is challenging. Therefore, we provide numerical verifications for two distributions used in the simulation and real data analyses.
Hereafter, we denote $N_1 = N_{a,b}^\obs = N(1- r_N^\mis)$ and $N_0 = N-N_{a,b}^\obs = N\cdot r_N^\mis$.  We assume that $\bF_N$ is a point mass distribution, and thus, $N_1$ and $r_N^\mis$ are deterministic.

\hspace{\fill}\\
\textbf{1. Normal distribution with both mean and variance unknown.}
Consider the case where $\vt_1,\ldots,\vt_{N} \sim \cN(\mu,\sigma^2)$ i.i.d. with $\mu,\ \sigma^2$ unknown.
The probability denisty is $g$.

\hspace{\fill}\\
\textbf{1.1. IRT$_\p$.}
We use the conjugate prior:
$\sigma^2 \sim \text{Inverse-Gamma}(\alpha_0,\beta_0)$, and 
$\mu\mid \sigma^2 \sim \cN(\mu_0,\sigma^2/\kappa_0)$.
Then the posterior distribution is 
\[\begin{split}
\sigma^2 \mid \bmvt_{1:N_1} &\sim \text{Inverse-Gamma}\left(\alpha_{N_1},
\beta_{N_1}\right),\\
\mu \mid \sigma^2,\bmvt_{1:N_1} &\sim \cN\left(\mu_{N_1},\sigma^2/\kappa_{N_1}\right),
\end{split}\]
where 
$$\alpha_{N_1}=\alpha_0+\frac{N_1}{2},\ 
\beta_{N_1}=\beta_0 + \frac{1}{2}\left[(N_1-1)S^2+\frac{N_1\kappa_0}{\kappa_{N_1}}(\bar{\bmvt}_{1:N_1}-\mu_0)^2\right],\
\mu_{N_1} = \frac{\kappa_0\mu_0+N_1\bar{\bmvt}_{1:N_1}}{\kappa_{N_1}},$$
and $S^2 = (N_1-1)^{-1}\sum_{i=1}^{N_1} (\vt_i-\bar{\bmvt}_{1:N_1})^2$, $\kappa_{N_1} = \kappa_0+N_1$.
Then the posterior predictive distribution is {a noncentral t-distribution:}
\[\begin{split}
\vt\mid \bmvt_{1:N_1} \sim t_{2\alpha_{N_1}}\left(\mu_{N_1},\frac{\beta_{N_1}}{\alpha_{N_1}}\left(1+\frac{1}{\kappa_{N_1}}\right)\right).
\end{split} 
\]
Let $\sigma^2_{N_1}=\frac{\beta_{N_1}}{\alpha_{N_1}}\left(1+\frac{1}{\kappa_{N_1}}\right)$, then 
\[\begin{split}
\hat{g}(\vt\mid\bmvt_{1:N_1}) = \frac{\Gamma(\alpha_{N_1}+1/2)}{\Gamma(\alpha_{N_1})\sqrt{2\alpha_{N_1}\pi\sigma^2_{N_1}}}
\left(1+\frac{(\vt-\mu_{N_1})^2}{2\alpha_{N_1}\sigma^2_{N_1}}\right)^{-\frac{2\alpha_{N_1}+1}{2}}.
\end{split}\]
In the following, we verify that 
\[\begin{split}
\bbE\left|\prod_{i=N_1+1}^{N}\frac{\hat{g}(\vt_i\mid\bmvt_{1:N_1})}{g(\vt_i)} - 1\right| = o(1)
\end{split}\]
holds for $r_N^\mis = O(N^{-1/2})$ through simulations.
For $N$ changing from 100 to 100000, let $r_N^\mis \propto N^{-\text{rate}}$ for $\text{rate} \in \{0.5,0.6,0.7\}$.
We generate $\vt_1,\ldots,\vt_N$ i.i.d. from the true distribution $\cN(0,1)$, {set $\alpha_0 = \beta_0 = \kappa_0 = 1$ and $\mu_0 = 0$,}
and compute
\[\begin{split}
\left|\prod_{i=N_1+1}^{N}\frac{\hat{g}(\vt_i\mid\bmvt_{1:N_1})}{g(\vt_i)} - 1\right|.
\end{split}\]
Repeat the above procedure 1000 times and the average of the obtained 1000 values is an approximation of 
\[\begin{split}
\bbE\left|\prod_{i=N_1+1}^{N}\frac{\hat{g}(\vt_i\mid\bmvt_{1:N_1})}{g(\vt_i)} - 1\right|.
\end{split}\]
Values of the approximation are shown in Fig. \ref{fig:par_normal}.
The results imply that when $r_N^\mis$ converges to 0 faster, $\bbE\left|\prod_{i=N_1+1}^{N}\frac{\hat{g}(\vt_i\mid\bmvt_{1:N_1})}{g(\vt_i)} - 1\right|$
 will converge to 0 more quickly.

\begin{figure}[h]
\centering
\includegraphics[width=0.7\linewidth]{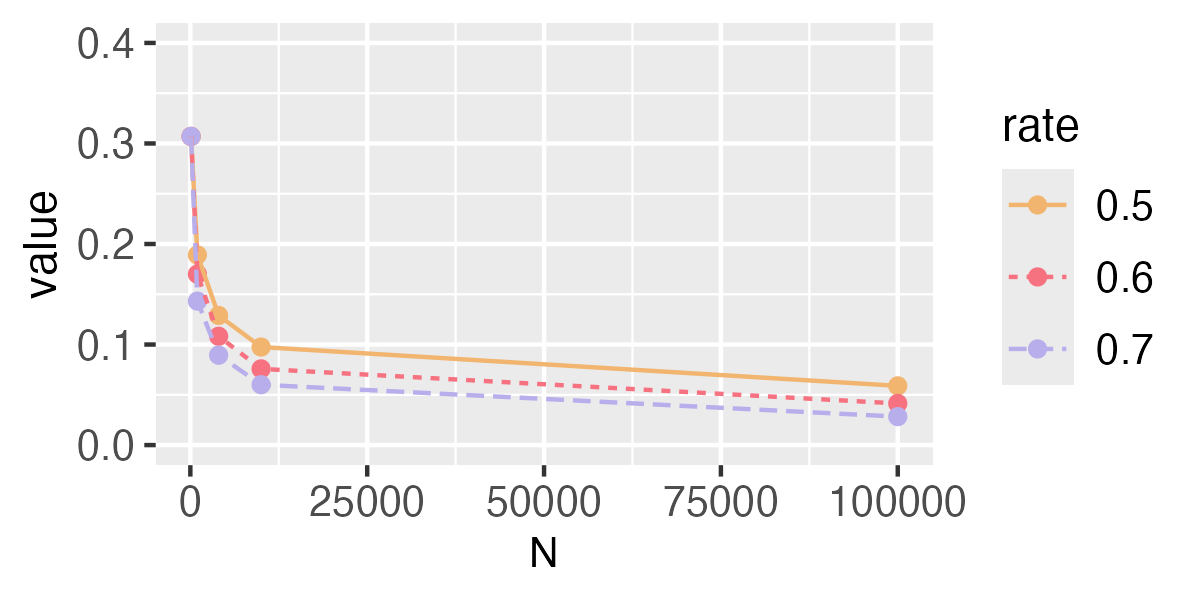}
\caption{Simulation results for IRT$_\p$ in the case of the normal distribution with both mean and variance unknown. }
\label{fig:par_normal}
\end{figure}

\hspace{\fill}\\
\textbf{1.2. IRT$_\s$.}
When $g(\cdot)$ is known to be continuous, let
\begin{equation}\label{eq:kernelfunction}
\hat{g}(\vt\mid \bmvt_{1:N_1}) = \frac{1}{N_1h_N}\sum_{j=1}^{N_1}K\left(\frac{\vt-\vt_j}{h_N}\right),
\end{equation}
where the kernel $K$ satisfies $\int K(u)du=1$, $\int u^2 K(u) du < \infty$, and $K(u) = K(-u)$.
IRT with imputation function \eqref{eq:kernelfunction} can be seen a smoothed version of IRT$_\e$ and we denote this method as IRT$_\s$.
Let $K$ be a Gaussian kernel and $g$ corresponds to normal distribution.
We choose $h_N\propto N_1^{-1/5}$ via a plug-in method, and verify that 
\[\begin{split}
\bbE\left|\prod_{i=N_1+1}^{N}\frac{\hat{g}(\vt_i\mid\bmvt_{1:N_1})}{g(\vt_i)} - 1\right| = o(1)
\end{split}\]
holds for $r_N^\mis = O(N^{-1/2})$ through simulations.
For $N$ changing from 100 to 100000, let $r_N^\mis \propto N^{-\text{rate}}$ for $\text{rate} \in \{0.5,0.6,0.7\}$.
We generate $\vt_1,\ldots,\vt_N$ i.i.d. from the true distribution $\cN(0,1)$, {set $\alpha_0 = \beta_0 = \kappa_0 = 1$ and $\mu_0 = 0$,}
and compute
\[\begin{split}
\left|\prod_{i=N_1+1}^{N}\frac{\hat{g}(\vt_i\mid\bmvt_{1:N_1})}{g(\vt_i)} - 1\right|.
\end{split}\]
Repeat the above procedure 1000 times and the average of the obtained 1000 values is an approximation of 
\[\begin{split}
\bbE\left|\prod_{i=N_1+1}^{N}\frac{\hat{g}(\vt_i\mid\bmvt_{1:N_1})}{g(\vt_i)} - 1\right|.
\end{split}\]
Values of the approximation are shown in Fig. \ref{fig:kernel_normal}.
The results imply that when $r_N^\mis$ converges to 0 faster, $\bbE\left|\prod_{i=N_1+1}^{N}\frac{\hat{g}(\vt_i\mid\bmvt_{1:N_1})}{g(\vt_i)} - 1\right|$
 will converge to 0 more quickly.

\begin{figure}[h]
\centering
\includegraphics[width=0.7\linewidth]{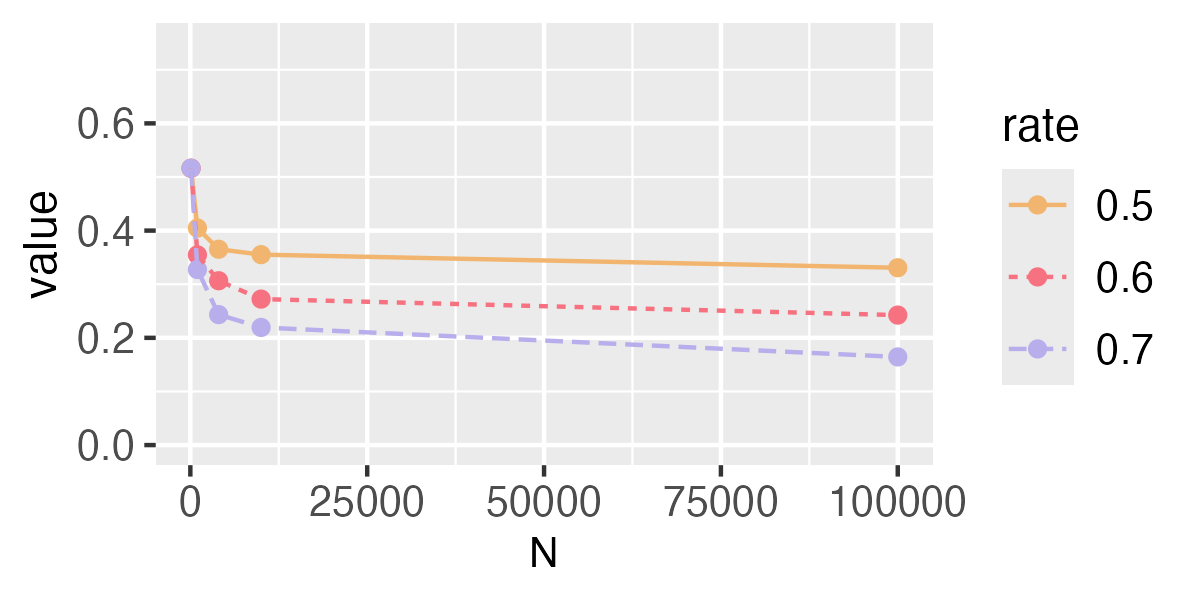}
\caption{Simulation results for IRT$_\s$ in the case of normal distribution with both mean and  variance unknown.}
\label{fig:kernel_normal}
\end{figure}

\hspace{\fill}\\
\textbf{2. Binomial distribution.}
Consider the case where $\vt_1,\ldots,\vt_{N} \sim \text{Binomial}(m,q)$ { with $m$ known} and $q$ unknown.
The corresponding probability mass is $p(\cdot)$.

\hspace{\fill}\\
\textbf{2.1. IRT$_\p$.}
We use prior $q \sim \text{Beta}(\alpha,\beta)$, where 
$\alpha,\beta > 0$. 
Then the posterior distribution is 
$q \mid \bmvt_{1:N_1} \sim \text{Beta}(\alpha_{N_1},\beta_{N_1})$, where 
\[\begin{split}
\alpha_{N_1} &= \alpha + N_1\cdot \bar{\bmvt}_{1:N_1},\ 
\beta_{N_1} = \beta + N_1(m-\bar{\bmvt}_{1:N_1}).
\end{split}\]
Then the posterior predictive distribution is 
\[\begin{split}
\vt\mid \bmvt_{1:N_1} \sim \text{Beta-Binomial}(m,\alpha_{N_1},\beta_{N_1}).
\end{split}\]
Denote the p.m.f. of this posterior predictive distribution as $\hat{p}(\vt \mid \bmvt_{1:N_1})$.
In the following, we assume that the sequence of $r_N^\mis$ is deterministic for $N_1 \sim \bF_N$ and verify that 
\[\begin{split}
\bbE\left|\prod_{i=N_1+1}^{N}\frac{\hat{p}(\vt_i\mid\bmvt_{1:N_1})}{p(\vt_i)} - 1\right| = o(1)
\end{split}\]
holds for $r_N^\mis = O(N^{-1/2})$ through simulations.
For $N$ changing from 100 to 100000,
let $r_N^\mis \propto N^{-\text{rate}}$ for $\text{rate} \in \{0.5,0.6,0.7\}$.
We generate $\vt_1,\ldots,\vt_N$ i.i.d. from the true distribution Binomial$(100,0.1)$, { set $\alpha = \beta = 1$,}
and compute
\[\begin{split}
\left|\prod_{i=N_1+1}^{N}\frac{\hat{p}(\vt_i\mid\bmvt_{1:N_1})}{p(\vt_i)} - 1\right|.
\end{split}\]
Repeat the above procedure 1000 times and the average of the obtained 1000 values is an approximation of 
\[\begin{split}
\bbE\left|\prod_{i=N_1+1}^{N}\frac{\hat{p}(\vt_i\mid\bmvt_{1:N_1})}{p(\vt_i)} - 1\right|.
\end{split}\]
Values of the approximation are shown in Fig. \ref{fig:par_binom}.
The results imply that when $r_N^\mis$ converges to 0 faster, $\bbE\left|\prod_{i=N_1+1}^{N}\frac{\hat{g}(\vt_i\mid\bmvt_{1:N_1})}{g(\vt_i)} - 1\right|$
 will converge to 0 more quickly.

\begin{figure}
\centering
\includegraphics[width=0.7\linewidth]{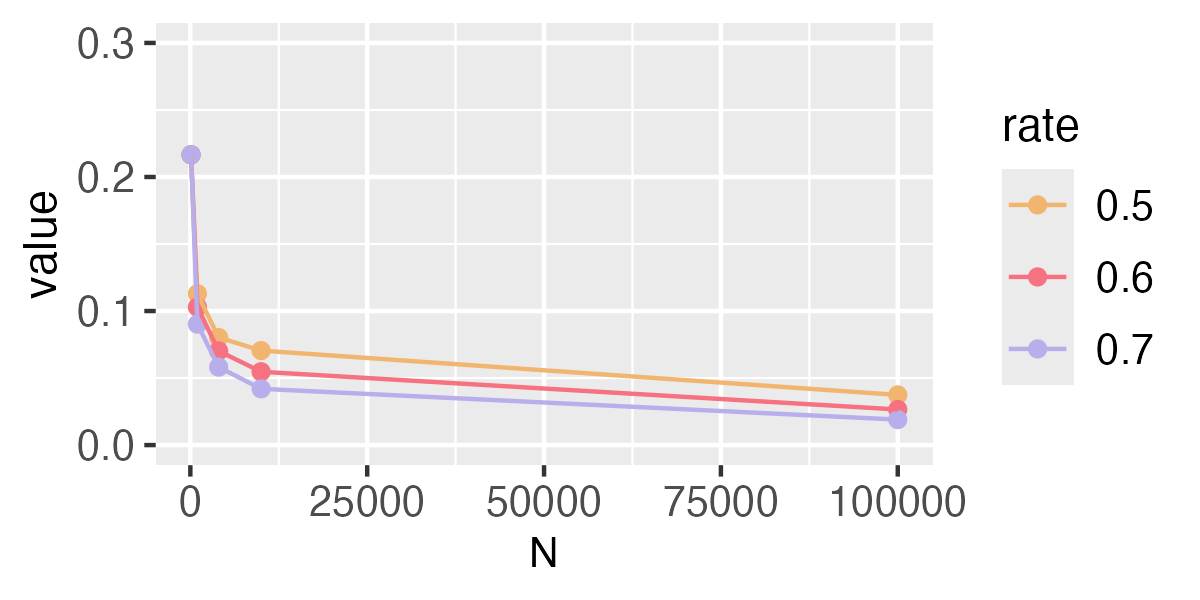}
\caption{Simulation results for IRT$_\p$ in the case of binomial distribution.}
\label{fig:par_binom}
\end{figure}

\hspace{\fill}\\
\textbf{2.2. IRT$_\e$.}
When $g(\cdot)$ corresponds to a discrete distribution, it takes the form of 
\begin{equation}\label{eq:discrete}
g(\vt) = \sum_{\psi \in \Psi} p(\psi)\delta(\vt-\psi),
\end{equation}
where $\Psi$ is the set of all the $\psi$ s.t. $g(\psi) > 0$.
Suppose we estimate $p(\cdot)$ with the empirical distribution 
\begin{equation}\label{eq:emp}
\hat{g}(\vt\mid \bmvt_{1:N_1}) = \frac{1}{N_1}\sum_{j=1}^{N_1}\delta(\vt - \vt_j).
\end{equation}
Then the empirical distribution can be written as 
\[\begin{split}
\hat{g}(\vt\mid \bmvt_{1:N_1}) = \sum_{\psi \in \Psi} \left(\frac{1}{N_1}\sum_{j=1}^{N_1} 1\{\vt_j = \psi\}\right)\delta(\vt-\psi),
\end{split}\]
i.e.,
\[\begin{split}
\hat{p}(\vt\mid \bmvt_{1:N_1}) = \frac{1}{N_1}\sum_{j=1}^{N_1} \bI\{\vt_j = \vt\}.
\end{split}\]
In the following, we verify that 
\[\begin{split}
\bbE\left|\prod_{i=N_1+1}^{N}\frac{\hat{p}(\vt_i\mid\bmvt_{1:N_1})}{p(\vt_i)} - 1\right| = o(1)
\end{split}\]
holds for $r_N^\mis = O(N^{-1/2})$ through simulations.
For $N$ changing from 100 to 100000,
let $r_N^\mis \propto N^{-\text{rate}}$ for $\text{rate} \in \{0.5,0.6,0.7\}$.
We generate $\vt_1,\ldots,\vt_N$ i.i.d. from the true distribution Binomial$(100,0.1)$, { set $\alpha = \beta = 1$,}
and compute
\[\begin{split}
\left|\prod_{i=N_1+1}^{N}\frac{\hat{p}(\vt_i\mid\bmvt_{1:N_1})}{p(\vt_i)} - 1\right|.
\end{split}\]
Repeat the above procedure 1000 times and the average of the obtained 1000 values is an approximation of 
\[\begin{split}
\bbE\left|\prod_{i=N_1+1}^{N}\frac{\hat{p}(\vt_i\mid\bmvt_{1:N_1})}{p(\vt_i)} - 1\right|.
\end{split}\]
Values of the approximation are shown in Fig. \ref{fig:emp_binom}.
The results imply that when $r_N^\mis$ converges to 0 faster, $\bbE\left|\prod_{i=N_1+1}^{N}\frac{\hat{g}(\vt_i\mid\bmvt_{1:N_1})}{g(\vt_i)} - 1\right|$
 will converge to 0 more quickly.

\begin{figure}[h]
\centering
\includegraphics[width=0.7\linewidth]{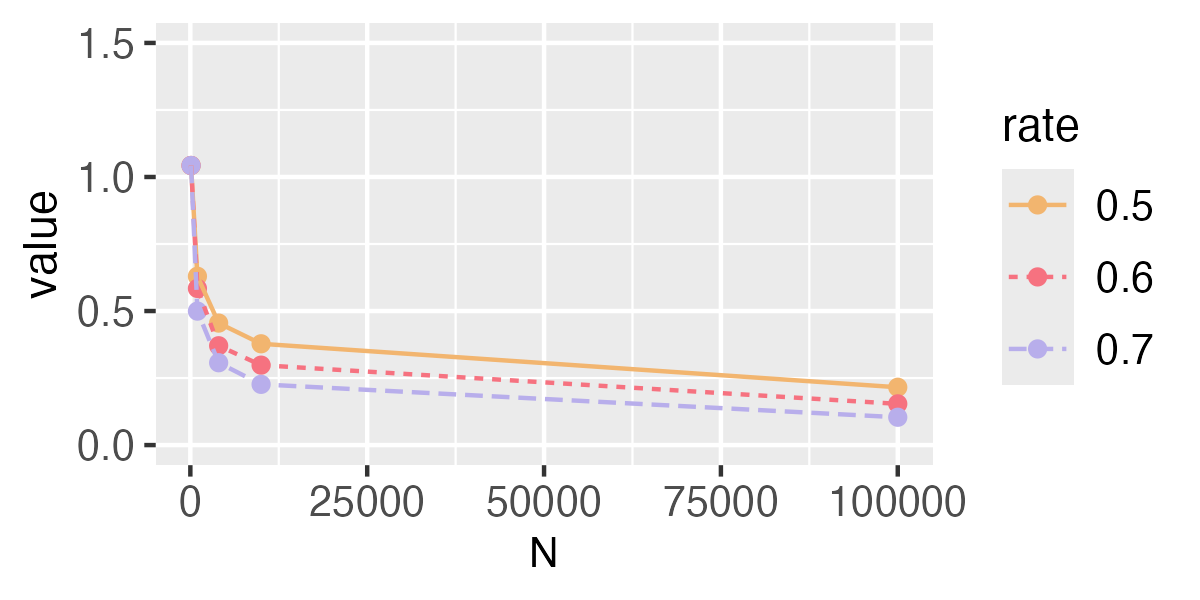}
\caption{Simulation results for IRT$_\e$ in the case of binomial distribution.}
\label{fig:emp_binom}
\end{figure}

\newpage
\section{Additional simulation results}
\label{sec:sim}
\setcounter{figure}{0}
\renewcommand{\thefigure}{B\arabic{figure}}

\subsection{Clustered Interference}

For simulations in Section 4.1, the type I error across 10 datasets and average power over these datasets with $Y_i(0) \sim \chi^2(4)$ and $t(4)$ are shown in Fig. \ref{fig:cluster_chisq4} and Fig. \ref{fig:cluster_t4}. 
In these scenarios, the prior for IRT$_p$ is still set to be a normal distribution.
Similar to the observations in Section 4.1, all methods control the type I error well and our IRT methods outperform CRTs and PNRTs with respect to the power when $Y_i(0) \sim \chi^2(4)$ or $t(4)$.
These results show that although the models used in IRT$_\p$ (i.e., normal) and IRT$_\e$ are different from the true distribution, their performances are robust.
Moreover, all IRTs under the estimated imputation distribution are close to IRT$_\o$.
Additionally, when $g$ corresponds to a continuous distribution, IRT$_\e$ and IRT$_\s$ are particularly close to each other in each case, which means that IRT$_\e$ can also work well when the real distribution is continuous.

\begin{figure}[ht]
\centering
\includegraphics[width=0.8\linewidth]{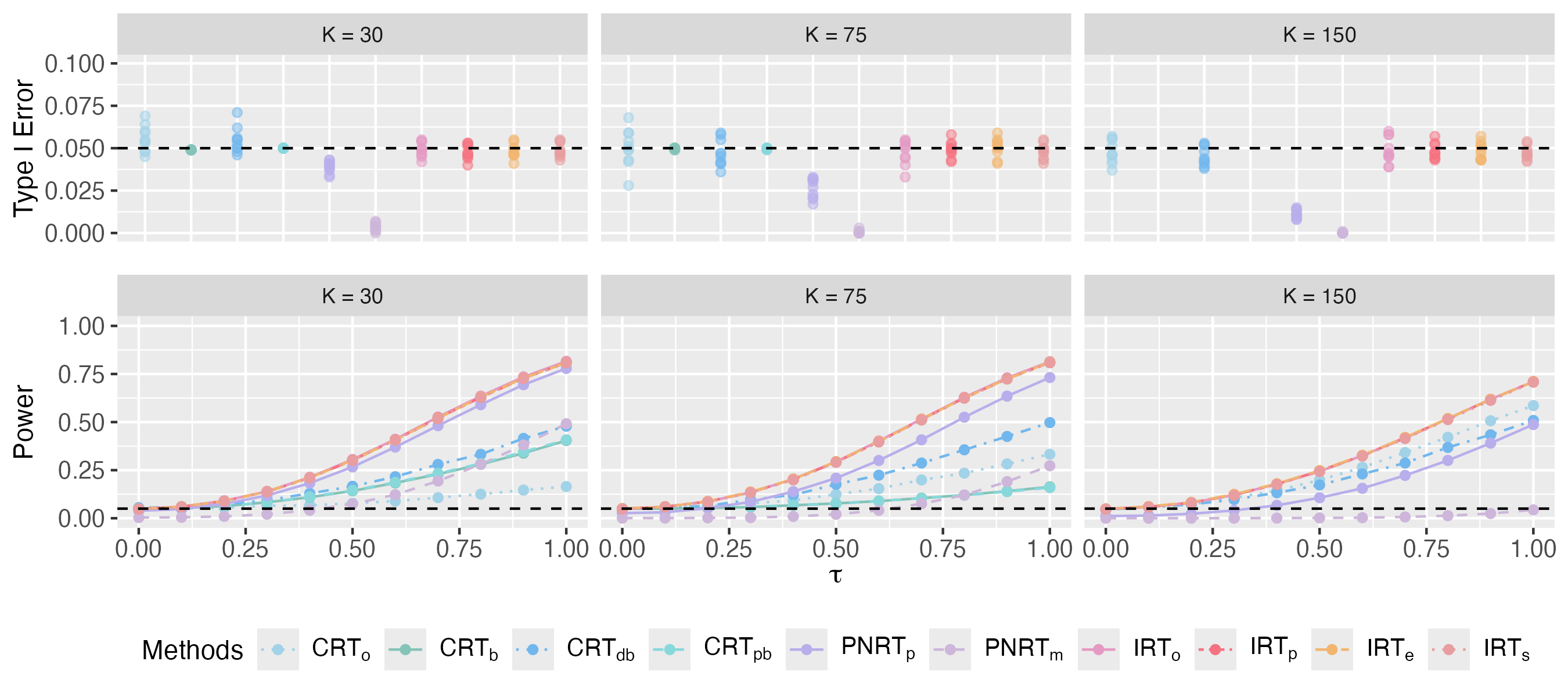}
\caption{Simulation results under clustered interference for $Y_i(0) \sim \chi^2(4)$. The type I error rate across 10 datasets and the average power over these datasets are visualized for 10 competing methods under different specifications of causal effect $\tau$ and number of clusters $K$. The horizontal dashed line indicates the significance level of $\alpha = 0.05$.}
\label{fig:cluster_chisq4}
\end{figure}

\begin{figure}[ht]
\centering
\includegraphics[width=0.8\linewidth]{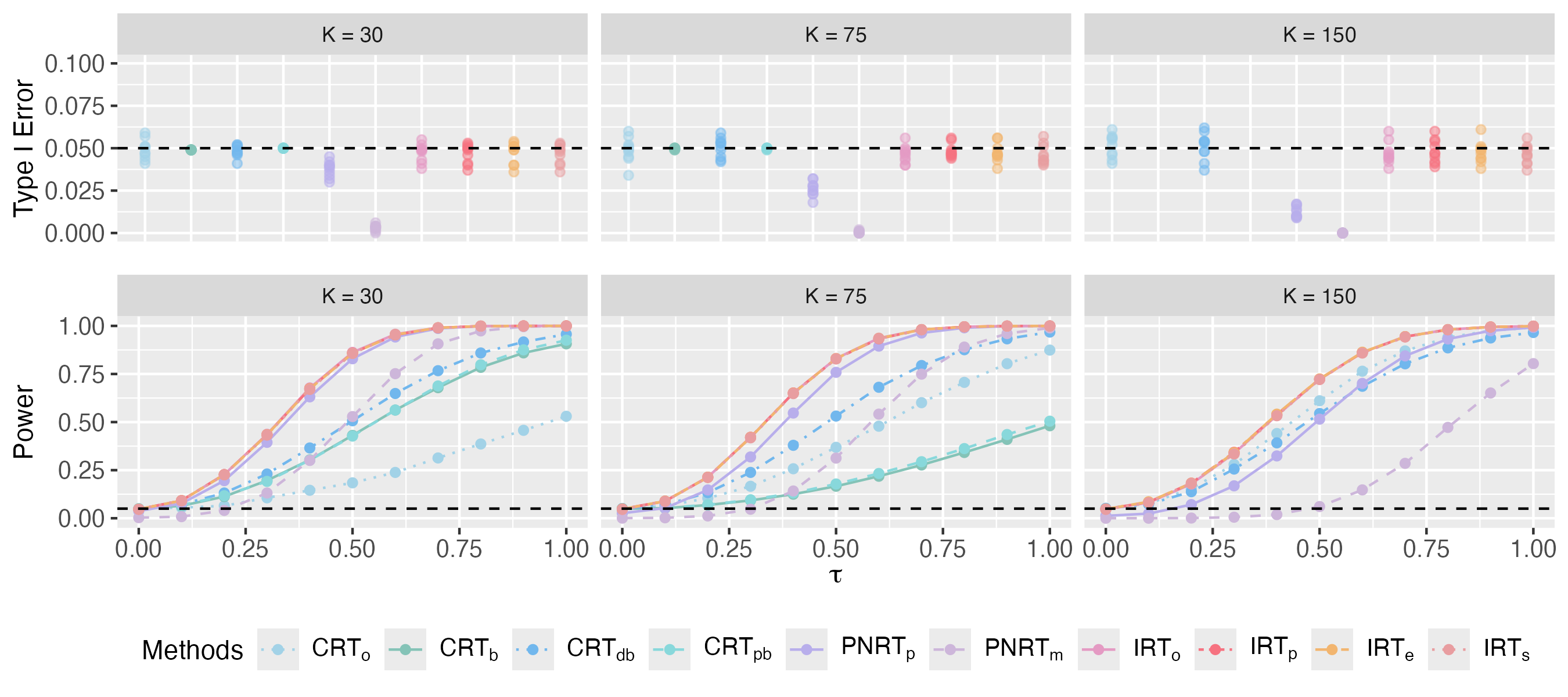}
\caption{Simulation results under clustered interference for $Y_i(0) \sim t(4)$. The type I error rate across 10 datasets and the average power over these datasets are visualized for 10 competing methods under different specifications of causal effect $\tau$ and number of clusters $K$. The horizontal dashed line indicates the significance level of $\alpha = 0.05$.}
\label{fig:cluster_t4}
\end{figure}

\FloatBarrier
\subsection{Spatial Interference}
For simulations in Section 4.2, the type I error across 10 datasets and the average power when $Y_i(0)\sim\chi^2(4)$ and  $t(4)$ are present in Fig. \ref{fig:spatial_power_chisq4} and Fig. \ref{fig:spatial_power_t4}.
The observations are similar to those in Section 4.2, and this implies that our IRT methods outperform CRTs and PNRTs, especially in the case when $p = 0.8$, where the missing rate of imputable outcomes is large.
Moreover, the results show that IRT methods are robust when model is mis-specified.

\begin{figure}[ht]
\centering
\includegraphics[width=0.8\linewidth]{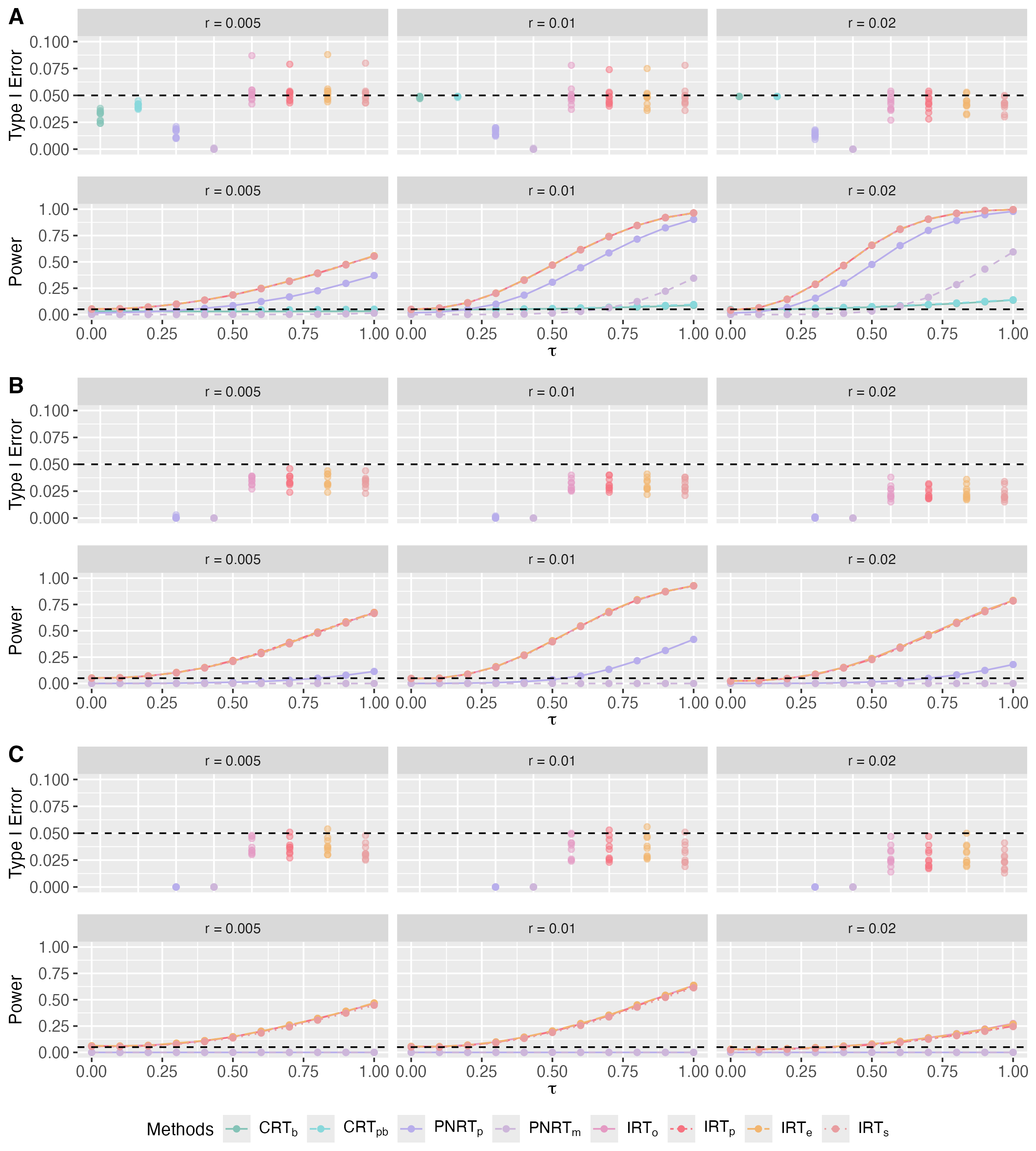}
\caption{Simulation results under spatial interference for $Y_i(0)\sim \chi^2(4)$: A, B, and C show the type I error rate and average power of 10 competing methods for $p=0.2, 0.5$, and $0.8$ respectively, under different specifications of casual effect $\tau$ and distance of interference $r$. The horizontal dashed line indicates the significance level of $\alpha = 0.05$.}
\label{fig:spatial_power_chisq4}
\end{figure}

\begin{figure}[ht]
\centering
\includegraphics[width=0.8\linewidth]{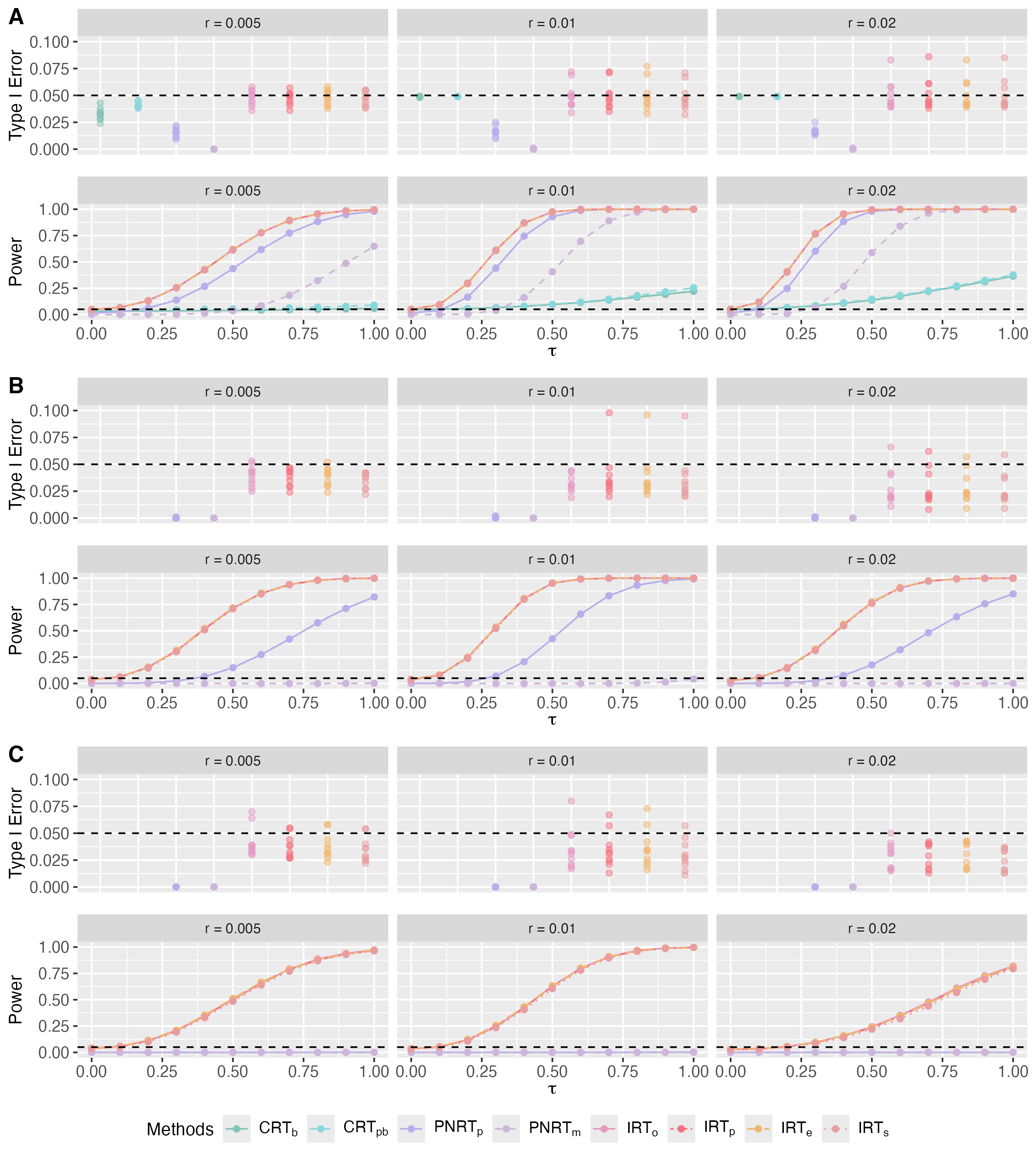}
\caption{Simulation results under spatial interference for $Y_i(0)\sim t(4)$: A, B, and C show the type I error rate and average power of 10 competing methods for $p=0.2, 0.5$, and $0.8$ respectively, under different specifications of casual effect $\tau$ and distance of interference $r$. The horizontal dashed line indicates the significance level of $\alpha = 0.05$.}
\label{fig:spatial_power_t4}
\end{figure}
\FloatBarrier


\end{document}